%% file: Karachentsev_n_en.tex
\documentclass[
aps,%
11pt,%
final,%
notitlepage,%
oneside,%
onecolumn,%
nobibnotes,%
nofootinbib,%
superscriptaddress,%
noshowpacs,%
centertags]%
{revtex4}
\input{sao_cmd_author.tex}

\begin{document}
\selectlanguage{english}


\title{Dwarf Galaxies in the Local Volume}

\author{\firstname{I.~D.}~\surname{Karachentsev}}
\email{ikar@sao.ru}
\author{\firstname{E.~I.}~\surname{Kaisina}}
\affiliation{\saoname}

\received{January 17, 2019}%
\revised{February 4, 2019}%
\accepted{February 4, 2019}%

\begin{abstract}
We review observational data about a sample of Local Volume
objects containing about 1000 galaxies within 11~Mpc of the Milky
Way. Dwarf galaxies with stellar masses $M_*/M_{\odot} < 9$~dex
make up 5/6 of the sample. Almost  40\% of them have their
distances measured with high precision using the Hubble Space
Telescope. Currently, the LV is the most representative and least
selection-affected sample of dwarf galaxies suitable for testing
the standard $\Lambda$CDM paradigm at the shortest cosmological
scales. We discuss the  H\,II properties of dwarf galaxies in
different environments and the star formation rates in these
systems as determined from  $FUV$\!- and H$\alpha$-survey data. We
also pay certain attention to the baryonic Tully--Fisher relation
for low-mass dwarf galaxies. We also point out that LV dwarfs are
important ``tracers'' for determining the total masses of nearby
groups and the nearest Virgo cluster.
\end{abstract}

\maketitle

\section{INTRODUCTION}

When compiling catalogs of galaxies astronomers usually limit the
sample by a certain limiting magnitude or a certain smallest
angular diameter of objects. Theoreticians modeling the evolution
of the large-scale structure of the Universe need volume limited
catalogs compare their computations with observations. Because of
the enormous variety of galaxies in terms of size (by a factor of
1000) and luminosity (by a factor of a million), magnitude- and
distance-limited samples have very few objects in common.

A distance-limited sample contains a much higher fraction of
low-luminosity objects (dwarfs) than usual catalogs. However, to
compile such a sample one needs accurate data about the individual
distances to galaxies. For distant galaxies their redshift $z$ is
quite a suitable distance indicator. At the same time, in the
nearby volume, where dwarfs systems can be discerned, Hubble
distances \mbox {$D=cz/H_0$} are not reliable because of the
peculiar velocities. Kraan-Korteweg and Tammann~\cite{kra1979:Karachentsev_n_en}
were the first to attempt compiling a catalog of nearby galaxies.
They composed a list of 179 galaxies across the entire sky except
for he virial domain of Virgo cluster with Local-Group reference
frame radial velocities \mbox {$V_{\rm LG}<500$}~km\,s$^{-1}$.

Compiling a representative sample of nearby galaxies became a
feasible task after the launch of the Hubble Space Telescope
(HST). In HST images taken in two colors galaxies are seen
resolved into stars, making it possible to determine the position
of the tip of the red-giant  branch~(TRGB) and use it to estimate
the distance to the galaxy with an accuracy of 5--10\%. This
method is applicable to practically all galaxies containing
stellar population older than 2~billion years. Exposures fitting
into one orbital period of the satellite allow TRGB distances to
be measured to galaxies located out to 11~Mpc from the Sun. The
volume of the region of this radius is usually referred to as the
Local Volume (LV). Our brief review focuses on the main properties
of low-luminosity  galaxies within $D=11$~Mpc.

As is evident from an analysis of the disk galaxies of
increasingly lower stellar mass, the signs of the spiral structure
in these objects begin to get fuzzy at masses  $M_*\leq
1\times10^9M_{\odot}$. Such  a mass is intermediate between those
of the two Milky-Way satellites---the Large and Small Magellanic
Clouds. At the same time, below this critical mass galaxies loose
signs of the central starlike core, which is usually associated
with a massive black holes. Galaxies with masses smaller than
$10^9M_{\odot}$ are usually called dwarf stellar systems. The
maximum amplitude of rotation velocity in these systems,
$V_m\leq50$~km\,s$^{-1}$, exceeds only slightly the characteristic
velocity of turbulent motions of gas clouds. In other words,
random chaotic events play an important part in the world of dwarf
galaxies.

Van~den~Bergh~\cite{vanB1959:Karachentsev_n_en} carried out a systematic search for
dwarf galaxies on \mbox {(POSS-I)} images. A total of 243 dwarf
galaxies got  \mbox {DDO designations}. Later,
Karachentseva~\cite{kara1968:Karachentsev_n_en} explored the same POSS-I images to
find 318 more dwarf galaxies (KDG) with  $2^{\rm m}$--$3^{\rm m}$
lower luminosities and surface brightness compared to DDO objects.
Karachentseva and Sharina~\cite{kara1988:Karachentsev_n_en} combined the above lists
with the results of the searches for dwarf galaxies performed by
other authors including the searches carried out in Southern-sky
images  (ESO/SERC), to compile the "Catalog of Low Surface
Brightness Dwarf galaxies'', which included 1500 objects
distributed across the entire sky.

After the release of the Second Palomar Observatory Sky Survey
(POSS-II) carried out on fine-grain plates Karachentseva and her
colleagues performed systematic searches and found 600 more dwarf
galaxies with typical surface brightness levels lower than
25$^{\rm
m}$/$\Box\arcsec$~\cite{kara1998:Karachentsev_n_en,kara1999:Karachentsev_n_en,kara2000:Karachentsev_n_en,kar2000:Karachentsev_n_en,huch2001a:Karachentsev_n_en}.
Follow-up 21-cm line radial-velocity measurements of these objects
showed that most of them are nearby dwarf
systems~\cite{huch2000:Karachentsev_n_en,huch2001b:Karachentsev_n_en,huch2003:Karachentsev_n_en}.

Many new dwarf galaxies were resolved into stars on images taken
with the  \mbox {6-m} telescope of the Special Astrophysical
Observatory of the Russian Academy of Sciences, making it possible
to estimate the distances to these galaxies from the luminosities
of their brightest blue and red supergiants~\mbox
{\cite{maka1997:Karachentsev_n_en,geo1997:Karachentsev_n_en,tik1998:Karachentsev_n_en,sha1999:Karachentsev_n_en}}. The researchers from
the Special Astrophysical Observatory of the Russian Academy of
Sciences found three new members of the Local group among many
nearby objects: LGS-3~\cite{kara1976:Karachentsev_n_en,mil2001:Karachentsev_n_en}, Cas~dSph, and
Peg~dSph~\cite{tik1999:Karachentsev_n_en}. Note, just for fun, that several dozen
dwarf galaxies found by Karachentseva, were later rediscovered and
renamed by other authors. The most recent example is the diffuse
dwarf system KKSG4~\cite{kar2000:Karachentsev_n_en}, which was rediscovered by
van-Dokkum et al.~\cite{vanD2018:Karachentsev_n_en}, which they named NGC\,1052-DF2.

Modern large-area CCD surveys: SDSS~\cite{aba2009:Karachentsev_n_en},
Pan-STARRS1~\cite{cha2016:Karachentsev_n_en}, and blind 21-cm surveys\linebreak
HIPASS~\cite{zwa2003:Karachentsev_n_en,kor2004:Karachentsev_n_en} and ALFALFA~\cite{gio2005:Karachentsev_n_en,hay2011:Karachentsev_n_en}
brought about the discovery of many gas-rich dwarf galaxies.
Dedicated CCD surveys of the vicinity of the neighboring Andromeda
galaxy M\,31~\cite{iba2007:Karachentsev_n_en}, Magellanic Clouds~\cite{drl2015:Karachentsev_n_en}, and
also the neighborhoods of other nearby massive galaxies---M\,81,
Cen\,A, M\,101,
M\,106~\cite{chi2009:Karachentsev_n_en,mul2017:Karachentsev_n_en,vanD2014:Karachentsev_n_en,kim2011:Karachentsev_n_en}---proved to be very
productive.

In the immediate vicinity of the Milky Way and the Andromeda
galaxy several dozen ultrafaint dwarf systems were found that
could be resolved into stars and whose sizes (about $10$~pc) and
stellar masses (about $10^5M_{\odot}$) are comparable with those
of globular clusters. Zwicky~\cite{zwi1957:Karachentsev_n_en} proposed to call them
``pygmy'' galaxies, but this term did not take hold in present-day
literature.

Note that besides large telescopes small instruments with
apertures of 1~m or less are also successfully used to search for
faint satellites of massive nearby
galaxies~\cite{mar2014:Karachentsev_n_en,kar2015a:Karachentsev_n_en,jav2016:Karachentsev_n_en}. Several dozen hours long
exposures on small telescopes made it possible to discover dwarf
satellites with surface brightness below~$26^{\rm
m}$/$\Box\arcsec$.

Dedicated efforts of different observing teams brought about a
rapid increase of the number of known dwarf galaxies in the LV. In
2004 ``A Catalog of Neighboring Galaxies'' contained 450 objects
within 10~Mpc~\cite{kar2004:Karachentsev_n_en}. A decade later, the ``Updated Nearby
Galaxy Catalog''~\cite{kar2013a:Karachentsev_n_en} already contained  869 galaxies
with estimated distances \mbox {$D<11$}~Mpc. As of now, the
regularly updated LV galaxy database~\cite{kai2012:Karachentsev_n_en} contains 1170
candidate members of the LV sample. Various data about these
galaxies are available at \mbox{\url{http://www.sao.ru/lv/lvgdb}}.
Dwarf galaxies with $M_*<9$~dex make up about 85\% of this sample.

\section{BASIC PROPERTIES OF THE SAMPLE OF GALAXIES IN THE LOCAL VOLUME}

We included a galaxy into the LV if its distance estimated using
some method did not exceed 11~Mpc. In the cases where only the
radial velocity of the galaxy was known it had to be $V_{\rm
LG}<600$~km\,s$^{-1}$ for the galaxy to be included into the LV
sample. Figure~1 shows the picture of the Hubble flow of galaxies
in the LV.

\begin{figure*}
\setcaptionmargin{5mm} \onelinecaptionstrue \captionstyle{normal}
\includegraphics[scale=1.1]{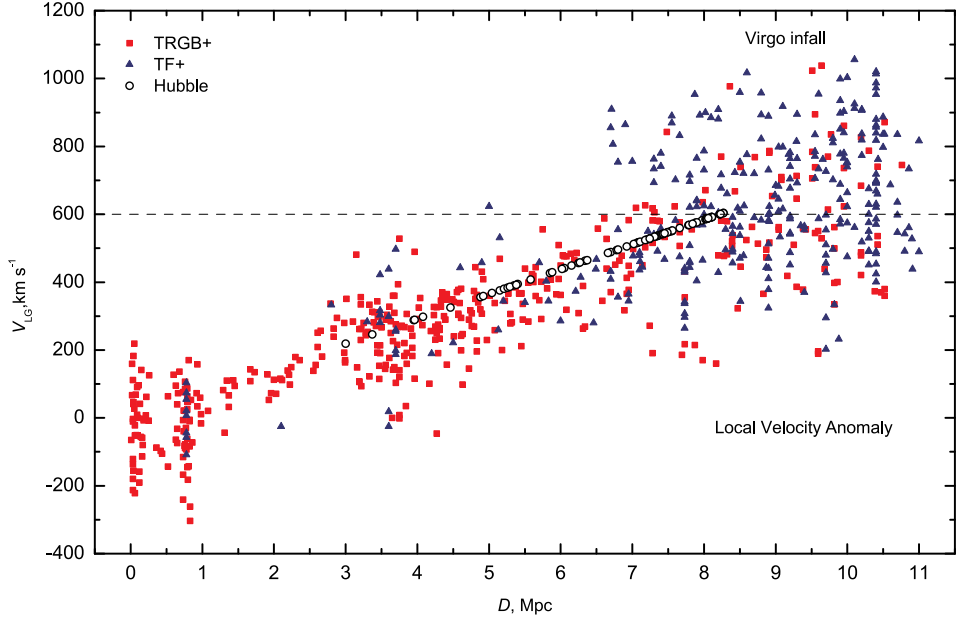}
\caption{The ``radial velocity--distance'' relation for LV
galaxies. The squares show the distance estimates obtained using
high-precision methods (TRGB etc.). The triangles show the
distance estimates obtained using less precise methods (TF etc.).
The open circles show the kinematic distances computed with the
adopted Hubble parameter of 73~km\,s$^{-1}$\,Mpc$^{-1}$. }
\end{figure*}

The distances $D$ of the galaxies are measured with respect to the
observer and the radial velocities $V_{\rm LG}$ are corrected for
the motion of the observer relative to the mass center of the
Local group~\cite{kar1996:Karachentsev_n_en}.

Galaxies with distance estimates obtained using different methods
are shown by differ symbols. About 90\% galaxies within  $D<6$~Mpc
have precise photometric distance estimates based on TRGB,
supernovae (SN), Cepheids (Cep), or surface-brightness
fluctuations~(SBF) and accurate to at least within 10\%. More than
half of all TRGB based measurements use HST observations performed
within the framework of programs proposed by the staff members of
the Special Astrophysical Observatory of the Russian Academy of
Sciences. Most of the distances near the far boundary of the LV
are estimated via the Tully--Fisher~(TF) method, fundamental
plane~(FP), from the planetary nebula luminosity function~(PNLF),
of based on the magnitudes of the brightest stars~(BS).  The
accuracy of these methods is not better than 20\%. We adopted
kinematic (Hubble) distances for some of the galaxies with \mbox
{$V_{\rm LG}<600$}~km\,s$^{-1}$. The dispersion of galaxy
velocities at small distances $D$ is due to the virial motions of
galaxies in groups. At $D>7$~Mpc effects of coherent streams begin
to show up: fall of galaxies toward the nearest massive attractor
in Virgo cluster and the motion of galaxies away from the center
of the Local cosmic void (the so-called phenomenon of  ``local
velocity anomaly''). It follows from these data that selecting LV
galaxies only by condition \mbox {$V_{\rm LG}<600$}~km\,s$^{-1}$
would introduce significant distortions in the local Hubble flow
pattern. On the average, the slope of the ``velocity--distance''
relation in Fig.~1 can be described quite well by the Hubble
parameter \mbox {$H_0=73$}~km\,s$^{-1}$\,Mpc$^{-1}$.

 \begin{figure*}
\setcaptionmargin{5mm} \onelinecaptionstrue \captionstyle{normal}
\includegraphics[scale=1.1, angle=90]{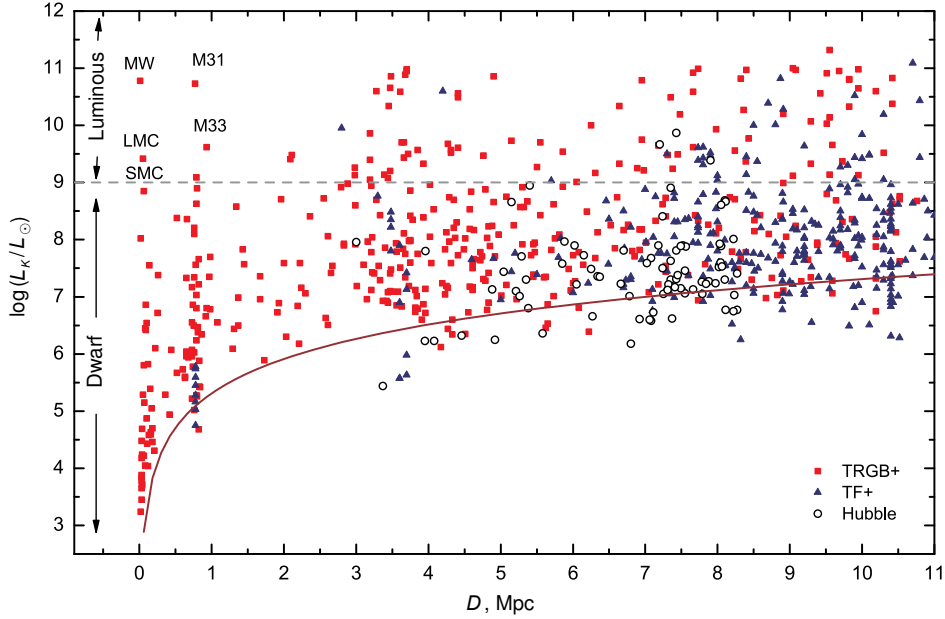}
\caption{The distribution of LV galaxies in the $K$-band
luminosity--distance plane. The symbol designations are the same
as in Fig.~1. The solid line corresponds to the apparent magnitude
$m_K=15^{\rm m}$. }
\end{figure*}

Figure~2 shows the distribution of LV galaxies in the $K$-band
luminosity--distance~$D$ plane. The designations of galaxies by
different symbols are the same\,as\,in\,Fig.\,1.
Given\,the\,\mbox{$M_*/L_K\!=\!1.0
M_{\odot}/L_{\odot}$}\,\cite{bel2003:Karachentsev_n_en}
mass-to-luminosity ratio, luminosity  $L_K$ can be used as a proxy
of the stellar mass of the galaxy. The da\-shed line separates
dwarf galaxies (\mbox{$M_*\!<\!1\times 10^9M_{\odot}$}) from
luminous galaxies. The solid line corresponds to the apparent
magnitude $m_K=15\fm0$, which approximately corresponds to the
photometric limit for all-sky searches for LV galaxies. The
distribution of galaxies relative to this curve is indicative of a
significant loss of ultrafaint objects with masses
$\log(M_*/M_{\odot})<6$ at the outskirts of the volume.

 \begin{figure*}
\setcaptionmargin{5mm} \onelinecaptionstrue \captionstyle{normal}
\includegraphics[scale=1.1, angle=90]{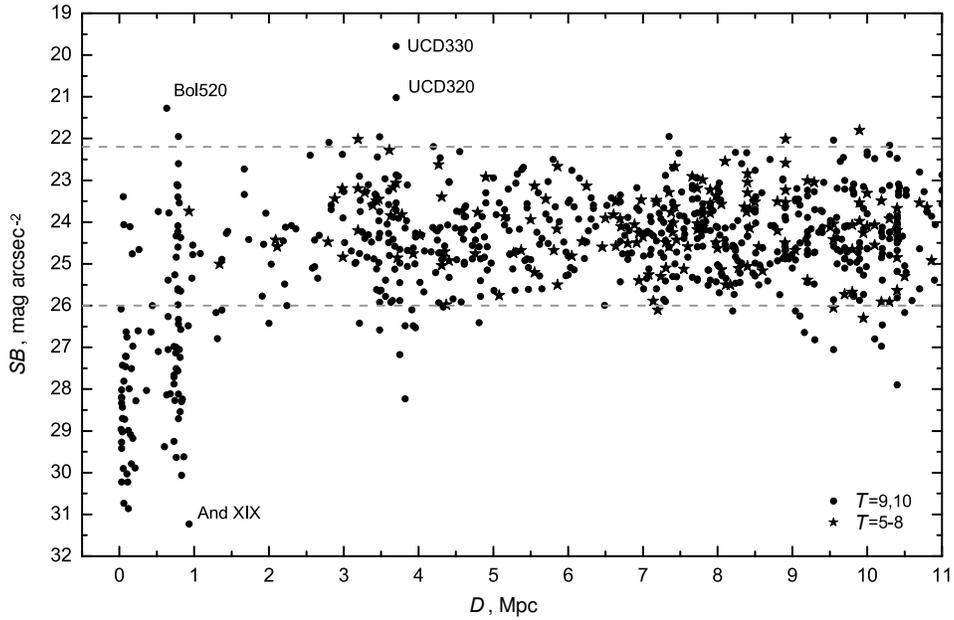}
\caption{The distance vs. average $B$-band surface brightness
distribution of LV galaxies. Late-type dwarf and spiral galaxies
are shown by different symbols.}
\end{figure*}

Figure~3 shows the distribution of average $B$-band surface
brightness  values of galaxies of our sample. The circles and
asterisks show late-type \mbox{($T = 9$ and $10$)} dwarf and
late-type ($T = 5$, $6$, $7$, and $8$) spiral galaxies,
respectively. Most of the galaxies have their average surface
brightness levels in the  $SB=[22;26]$~m/$\Box\arcsec$ interval,
which corresponds to 100\% and 3\% moonless night-sky brightness
level, respectively. The broadest surface-brightness interval,
\mbox {$SB=[21;30]$}~m/$\Box\arcsec$, is spanned by the satellites
of the Milky Way and Andromeda galaxy. Most of these galaxies were
discovered only when they were resolved into individual stars. We
can now extrapolate the situation in the Local group to the entire
LV assuming that about half of all very low surface-brightness
dwarf galaxies are lost in distant parts of the LV. Some
ultracompact dwarf galaxies like UCD\,320, UCD\,330 can also be
missed because of their almost starlike appearance.

\section{SPACE DISTRIBUTION OF NEARBY DWARFS}

About 52\% of dwarf galaxies reside in regions of gravitational
influence of luminous galaxies like the Milky Way and M31 and form
groups with sizes of about 200--300~kpc around these large
galaxies. The remaining nearby dwarfs are distributed among groups
clustering into diffuse filamentary and flat structures. On the
whole, dwarf and normal galaxies concentrate toward the plane of
the Local supercluster centered on Virgo cluster of galaxies.

 \begin{figure*}
\setcaptionmargin{5mm} \onelinecaptionstrue \captionstyle{normal}
\hbox{
\includegraphics[scale=0.63]{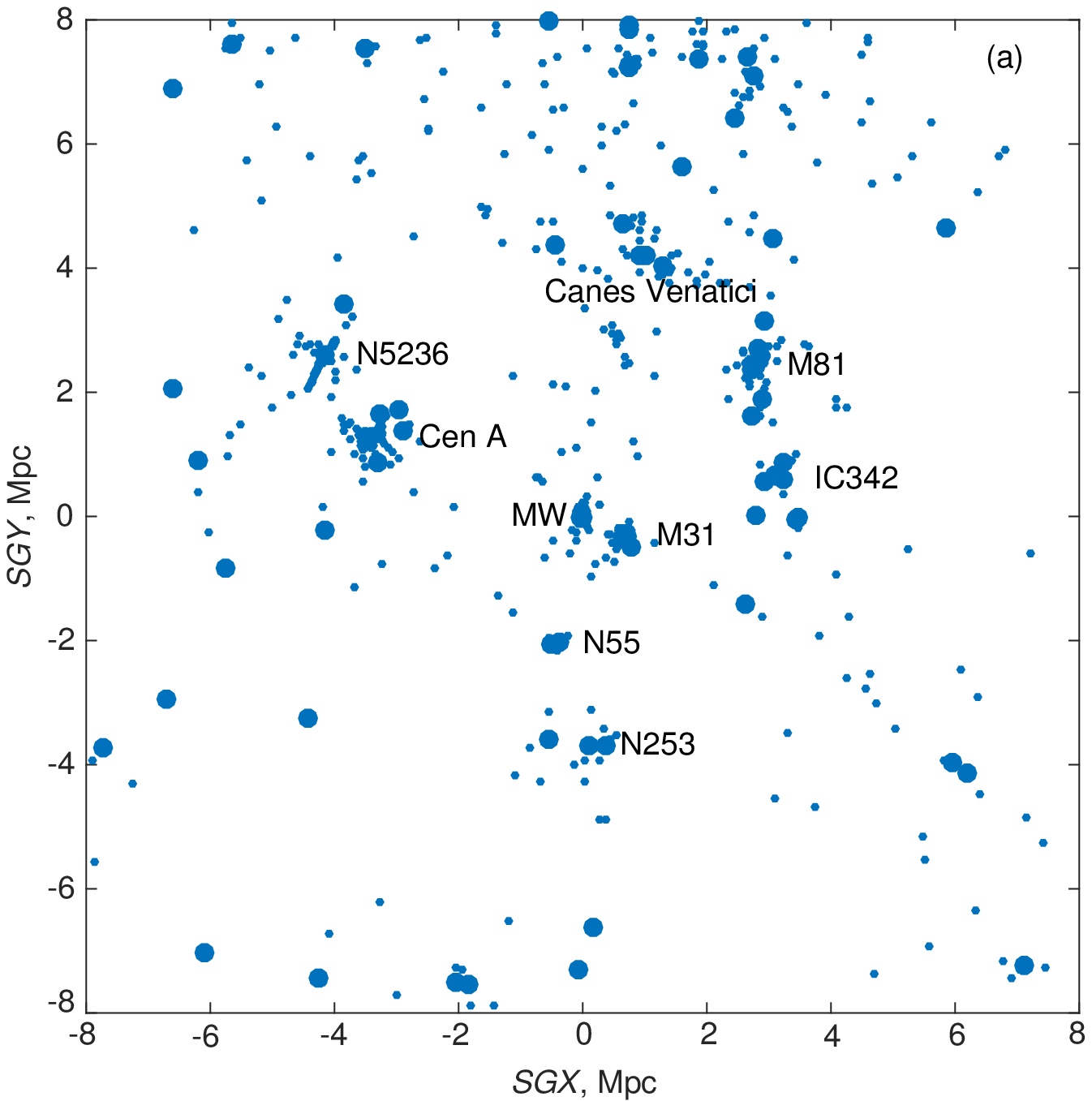}
\includegraphics[scale=0.63]{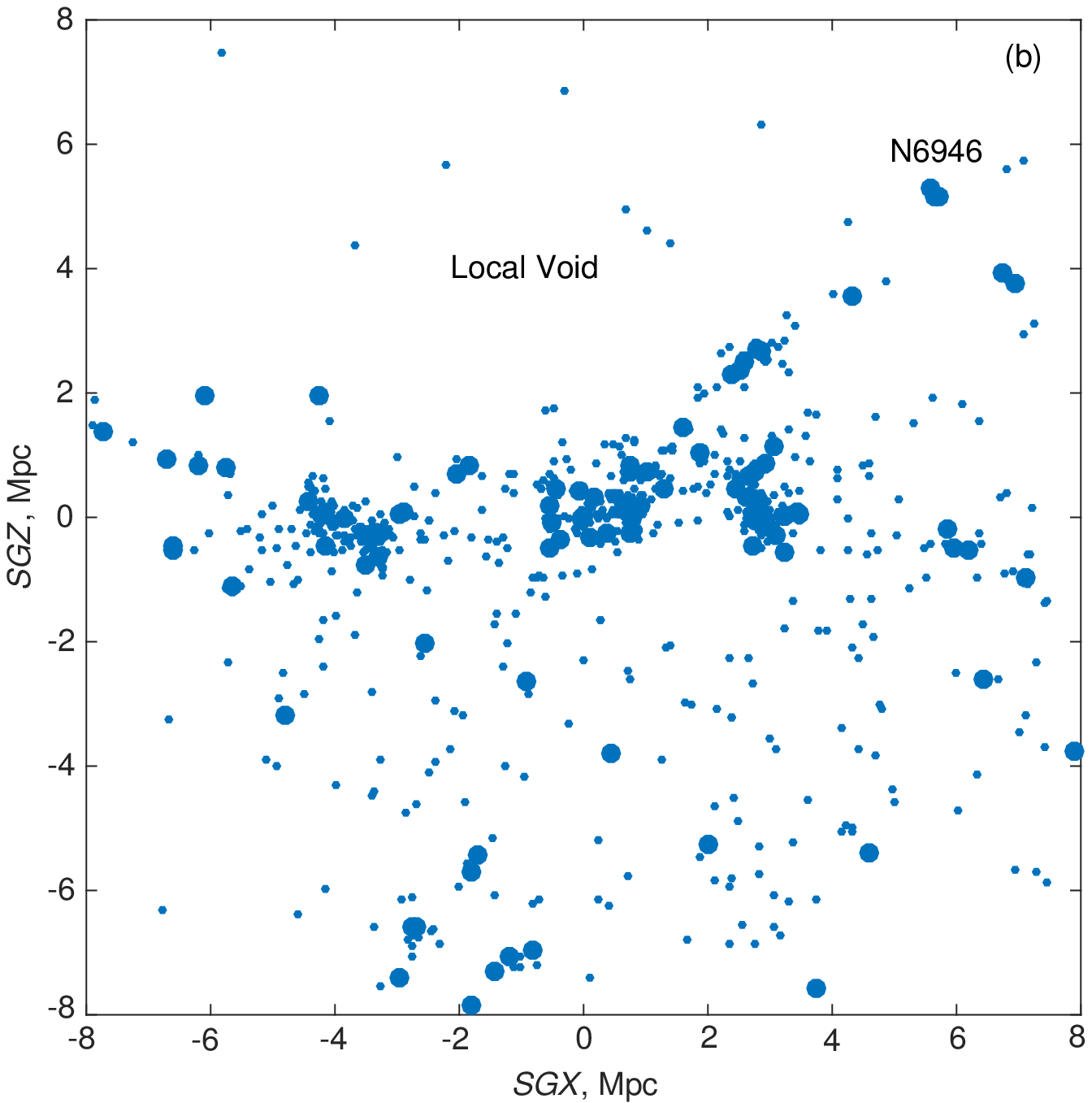}}
\caption{Distribution of nearby galaxies with distances $D<8$~Mpc
in supergalactic coordinates: (a)---view of galaxies in the plane
of the Local supercluster and located within 2~Mpc from this plane
(b)---edge-on view of the Local ``pancake''.}
\end{figure*}

\begin{table}[]
\setcaptionmargin{0mm} \onelinecaptionstrue \captionstyle{normal}
\caption{\label{MD:Karachentsev_n_en} Luminous galaxies in the LV}
\begin{tabular}{l|c|c|c|c}
\hline
Galaxy & $D$, & $V_{\rm LG}$,    & $\log M_*$,   & $\log M_{\rm tot}$,\\
                        & Mpc  & km\,s$^{-1}$ & [$M_{\odot}$] & [$M_{\odot}$]\\
\hline
 \multicolumn{1}{c|}{(1)} & (2) & (3) & (4) & (5)\\
\hline
Milky Way  & 0.01 & $-65$ & 10.70 & 12.07\\
M\,31      & 0.77 & $-29$ & 10.79 & 12.23\\
NGC5\,128  & 3.68 & 310   & 10.89 & 12.89\\
M\,81      & 3.70 & 104   & 10.95 & 12.69\\
NGC\,253   & 3.70 & 276   & 10.98 & 12.18\\
NGC\,4826  & 4.41 & 365   & 10.49 & 10.78\\
NGC\,4736  & 4.41 & 352   & 10.56 & 12.43\\
NGC\,5236  & 4.90 & 307   & 10.86 & 12.02\\
M\,101     & 6.95 & 378   & 10.79 & 12.17\\
NGC\,4258  & 7.66 & 506   & 10.92 & 12.50\\
NGC\,3627  & 8.32 & 579   & 10.82 & 12.16\\
M\,51      & 8.40 & 538   & 10.97 & 11.78\\
NGC\,2903  & 8.87 & 443   & 10.82 & 11.68\\
NGC\,5055  & 9.04 & 562   & 11.00 & 12.49\\
NGC\,4594  & 9.55 & 894   & 11.30 & 13.45\\
NGC\,6744  & 9.51 & 706   & 10.91 & 11.72\\
NGC\,3115  & 9.68 & 439   & 10.95 & 12.54\\
NGC\,2683  & 9.82 & 334   & 10.81 & 12.13\\
NGC\,891   & 9.95 & 736   & 10.98 & 11.90\\
NGC\,628   & 10.2 & 827   & 10.60 & 11.66\\
NGC\,3379  & 11.0 & 774   & 10.92 & 13.23\\
\hline
IC\,342    & 3.28 & 244   & 10.60 & 12.26\\
Maffei\,2  & 5.73 & 214   & 10.92 & 12.41\\
NGC\,6946  & 7.73 & 355   & 10.99 & 11.94\\
\hline
\end{tabular}
\end{table}

The two panels in Fig.~4 show the distribution of galaxies in the
nearby part of the LV within the radius $D=8$~Mpc. The left panel
shows the distribution of galaxies in the plane of the Local
supercluster without objects at high latitudes \mbox
{$|SGZ|>2$}~Mpc. Massive and dwarf galaxies are shown by circles
of different size. Besides well-defined compact groups around
M\,81 and Cen\,A, a diffuse Canes Venatici cloud can be seen at
the top of the figure, which consists mostly of dwarf galaxies. As
is evident from the figure, groups of galaxies are not distributed
randomly, but associate into filaments. Dwarf galaxies inside
groups are often arranged into flat structures with signs of
coherent motions~\mbox {\cite{iba2013:Karachentsev_n_en,mul2018:Karachentsev_n_en}}.

In the edge-on view of the Local ``wall''  (Fig.~4b) a large empty
region can be seen toward the north pole of the supercluster at
$|SGZ|>2$~Mpc populated by only a few dwarf galaxies. According to
Tully~\cite{tul1988:Karachentsev_n_en}, this Local cosmic void extends for more than
20~Mpc. An analysis of the radial velocities of galaxies shows
that the entire flat structure at $|SGZ|<2$~Mpc moves toward
``$-SGZ$'' from the center of the Local void at a velocity of
about 280~km\,s$^{-1}$~\cite{tul2008:Karachentsev_n_en}.

Dwarf galaxies can be used as ``tracers'' to determine the mass of
the central luminous galaxy. We assume that satellites move in
randomly oriented orbits with an average eccentricity of
 $\langle e^2\rangle=1/2$ to obtain the following estimate for the total mass:
  \begin{equation}
  M_T=16\pi G^{-1}\langle\Delta V^2 R_p\rangle,
  \end{equation}
where $\Delta V$ and $R_p$ are the difference of the radial
velocities of the satellite and the main galaxy and its projected
separation from the main galaxy, respectively, and $G$ is the
gravitational constant.

Inside the LV at Galactic latitudes \mbox{$|b|>15\degr$} there are
a total of 21 galaxies with stellar masses
$\log(M_*/M_{\odot})\geq10.5$.  The data about these galaxies are
listed in Table~1:~the first column gives the name of the galaxy;
the second and third columns, its distance and radial velocity,
respectively; the fourth and the fifth columns, the magnitude and
total (orbital) mass of the galaxy\footnote{For completeness, the
parameters of other three massive galaxies are given, which are
located at low Galactic latitudes and whose data are not quite
reliable because of strong extinction.}.  About half of the dwarf
population of the LV are located inside the regions of influence
of these giant galaxies. It follows from the data presented here
that the typical total-to-stellar mass ratio is $M_T/M_*\simeq32$.
A detailed analysis of the statistics of the $M_T/M_*$ ratios
galaxies of different morphological types can be found
in~\cite{kar2014a:Karachentsev_n_en,kar2014b:Karachentsev_n_en}.

Note that the dwarf galaxies in groups (suites) around massive
galaxies are mostly of the dSph type with almost total lack of gas
and young stellar population. The dwarf galaxy population of the
general field, on the contrary, is dominated by irregular~(dIr)
and blue compact galaxies~(BCD), in the stage of ongoing star
formation. Dwarf galaxies with their ``shallow'' potential wells
are believed to easily loose gas while moving through the halo of
their massive neighbor thereby exhausting the potential for
further star formation. However, recently,  ``dead'' spheroidal
galaxies KKR\,25~\mbox {\cite{kar2001:Karachentsev_n_en,mak2012:Karachentsev_n_en}},
Apples\,1~\cite{pas2005:Karachentsev_n_en}, KKs\,3~\cite{kar2015b:Karachentsev_n_en}, and
KK\,258~\cite{kar2014c:Karachentsev_n_en} have been found located quite far from
their nearest neighbors. The cause of their suppressed activity so
far remains unclear.

\section{H\,I-PROPERTIES OF DWARF GALAXIES IN DIFFERENT ENVIRONMENTS}

The morphology of dwarf galaxies is to a significant degree
determined by the availability of large reserves of gas, which
determine the possibility of further star formation detected by
H$\alpha$ emission flux or in far ultraviolet.

Table~2 gives a clue as to extent to which LV dwarfs have been
studied in the HI 21-cm line and in other parts of the
electromagnetic spectrum. The top row of the table gives the total
number of galaxies in the LV and separately the number of  dIr
\mbox {$(T= 10)$},  Im+BCD $(T= 9)$, and
 dSph+dE-type $(T < 0)$ dwarf galaxies.
Objects located in the zone of strong Galactic extinction
$(A_B>3\fm0)$, and those located outside  \mbox{$D=11.0$~Mpc} are
excluded. The second and third rows of the table give the numbers
of galaxies of the above types that were observed and detected in
the H\,I line. As is evident from the Table, H\,I-surveys cover
75\% of all LV galaxies. The detection levels are equal to 90\%
and 94\% for dIr and Im+BCD type galaxies and only 16\% for
dSph+dE type galaxies. The small observed number of undetected
irregular dwarfs and the small number detected spheroidal dwarfs
is due to the presence of the population of dwarf  ``transient''
systems (Tr) with morphology intermediate between dIr and dSph.
Another example of such objects, which cannot be clearly
classified, is LGS-3---a peripheral satellite of the Andromeda
galaxy.

 \begin{table}[]
 \setcaptionmargin{0mm} \onelinecaptionstrue \captionstyle{normal}
 \caption{Number of the LV galaxies observed and detected in H\,I, FUV, H$\alpha$}
\begin{tabular}{l|c|c|c|c}
 \hline
 Sample numbers & All types & dIr  & Im+BCD & dSph+dE \\
\hline
All LV galaxies       & 1072 & 404  & 152 & 346   \\
Observed in H\,I      & 806  & 335  & 132 & 180   \\
Detected in H\,I      & 596  & 300  & 124 & ~13   \\
Observed in $FUV$     & 914  & 351  & 123 & 294   \\
Detected in $FUV$     & 657  & 308  & 122 & ~81   \\
Observed in H$\alpha$ & 654  & 261  & 118 & 122   \\
Detected in H$\alpha$ & 532  & 221  & 112 & ~50   \\
\hline
   \end{tabular}
   \end{table}

The hydrogen mass of a late-type galaxy determined by the formula
    \begin{equation}
    (M_{\rm H\,I}/M_{\odot})=2.36\times10^5 D^2 F_{\rm H\,I},
    \end{equation}
where $F_{\rm H\,I}$ is the  H\,I line flux in Jy\,km\,s$^{-1}$,
correlates quite closely with the stellar mass of the galaxy.

\begin{figure*}
\setcaptionmargin{5mm} \onelinecaptionstrue \captionstyle{normal}
\includegraphics[scale=1.1]{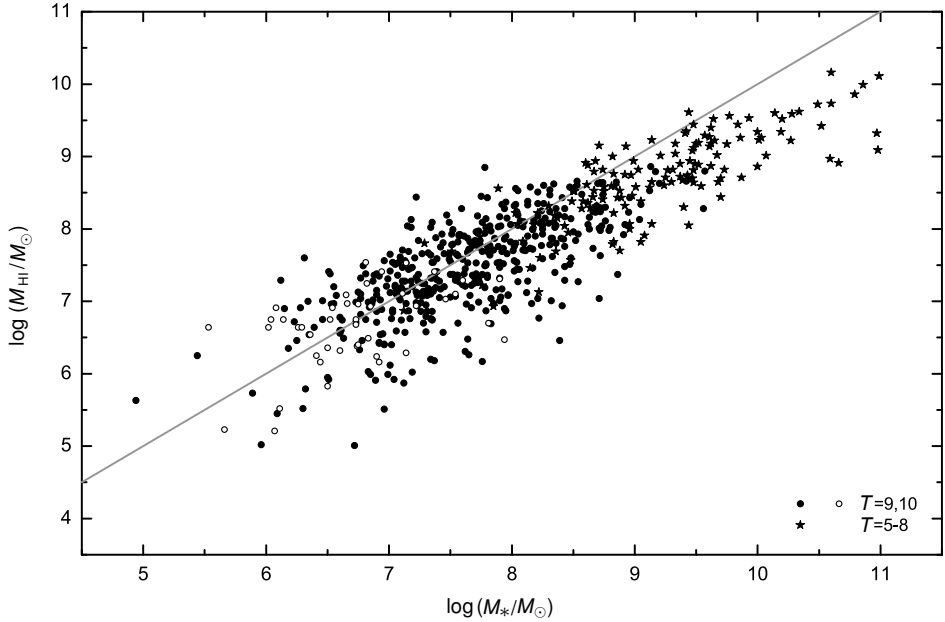}
\caption{Distribution of late-type $(T = 5$--$10)$ LV galaxies by
hydrogen and stellar masses. The open circles show the galaxies
with the upper H\,I-flux limit. The straight line corresponds to
the case of equal hydrogen and stellar masses.}
 \end{figure*}

Figure~5 reproduces the distribution of the number of late-type
dwarf (circles) and spiral (asterisks) LV galaxies by their
hydrogen and stellar masses. The straight line corresponds to the
condition $M_{\rm H\,I}=M_*$. The galaxies with upper $F_{\rm
H\,I}$ flux limits are shown by open circles. These data imply the
existence of many dwarf systems where the hydrogen mass has not
yet been transformed into stellar mass. Such cases occur most
often among the least luminous galaxies. On the contrary, disks of
luminous spiral galaxies mostly consist of the stellar component,
i.e., they show a more advanced evolutionary stage.

We already pointed out above that the closer the dwarf galaxy is
to its massive neighbor, the lower is its gas content. The effect
of the ambient density on the relative content of hydrogen mass in
LV galaxies was analyzed in detail in~\cite{kar2018a:Karachentsev_n_en}. To estimate
the ambient density, the above authors used three different
dimensionless parameters: $\Theta_1$, the relative local density
created by the most important neighbor in terms of mass and
distance; $\Theta_5$, the relative density created by the five
most important neighbors,  and $\Theta_j$, the relative
$K$-luminosity density inside the sphere of radius 1~Mpc in the
units of the average global luminosity density. All three
parameters are expressed in logarithmic scale.

 \begin{figure*}
\setcaptionmargin{0mm} \onelinecaptionstrue \captionstyle{normal}
\includegraphics[scale=1.4]{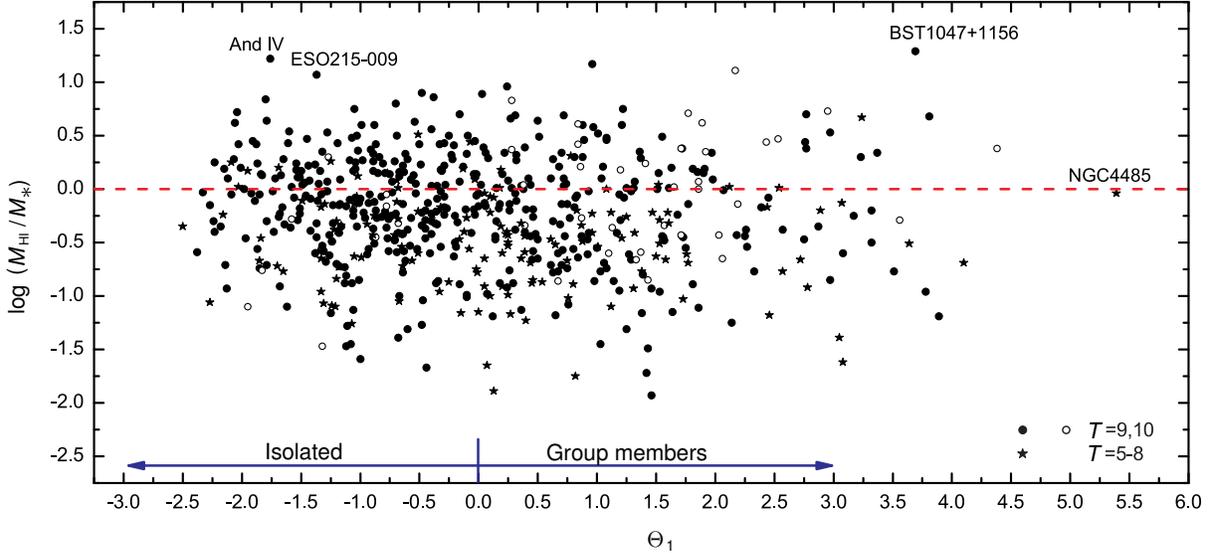}
\caption{Hydrogen-to-stellar mass ratio in late-type dwarf $(T =
9,~10)$ and spiral $(T = 5$--$8)$ galaxies in different
environments. Galaxies with upper  H\,I-flux limits are shown by
open circles.}
  \end{figure*}

Figure~6 shows the behavior of the  $M_{\rm H\,I}/M_*$ ratio for
late-type dwarf and spiral galaxies depending on tidal index
$\Theta_1$. Objects where only the upper  $F_{\rm H\,I}$ flux
limit is measured are shown by open circles. The domain of
negative $\Theta_1$ corresponds to isolated galaxies and the
right-hand side, where $\Theta_1>0$, is populated by satellites of
primary galaxies in groups. The scatter of $M_{\rm HI}/M_*$ ratios
is rather large and amounts to three orders of magnitude. In
extreme cases---BST\,1047+1156, And-IV and
\mbox{ESO\,215-009}---the hydrogen mass of the dwarf galaxy is one
order of magnitude higher than its  stellar mass. The average
$M_{\rm H\,I}/M_*$ ratio decreases slightly from isolated galaxies
to group members. This tendency becomes more pronounced if we take
into account the fact that the  $\Theta_1>0$ domain contains many
spheroidal dwarfs, which were not detected in the H\,I line. A
detailed analysis of these data is difficult to perform because of
the low resolution of the radio telescopes: the H\,I flux of many
dwarfs in groups can hardly been distinguished against the
background flux of the more massive neighbor.

The small drift of  $M_{\rm H\,I}/M_*$ on $\Theta_1$ indicates
that the hydrogen-to-stellar mass ratio in late- type galaxies is
primarily determined by their internal processes and only secondly
to ambient density.

\section{STAR FORMATION RATE IN DWARFS IN GROUPS AND IN THE FIELD}

The far ultraviolet survey  of galaxies carried out with
GALEX~\cite{gil2007:Karachentsev_n_en} orbital telescope detected
$FUV$\mbox{-}flu\-xes in a large number of galaxies located
outside the Milky Way zone. According to~\cite{lee2009:Karachentsev_n_en,lee2011:Karachentsev_n_en},
the integrated star-formation rate in the units of
$(M_{\odot}\,{\rm yr}^{-1}$) can be written as
     \begin{equation}
     \log(SFR)=2.78+2\log D-0.4 m^c_{FUV},
     \end{equation}
where $m^c_{FUV}$ is the apparent {$FUV$-band} magnitude of the
galaxy corrected for Galactic and internal extinction. The main
contributors to the  \mbox {$FUV$-flux} are young stars with
characteristic age of about 100~Myr.

It follows from Table~2 that GALEX survey area covers 85\% of LV
galaxies and 72\% them were actually detected in the $FUV$ band.
The fraction of detected galaxies among dIr and Im+BCD-type dwarfs
was equal to 88\% and 99\%, respectively, whereas only 28\% of
objects among  dSph+dE-type galaxies exhibited appreciable  $FUV$
flux. Note that $FUV$-band fluxes of some of the spheroidal
galaxies with large angular sizes may be false because of
projected background sources.

\begin{figure*}
  \setcaptionmargin{5mm} \onelinecaptionstrue \captionstyle{normal}
   \includegraphics[scale=1]{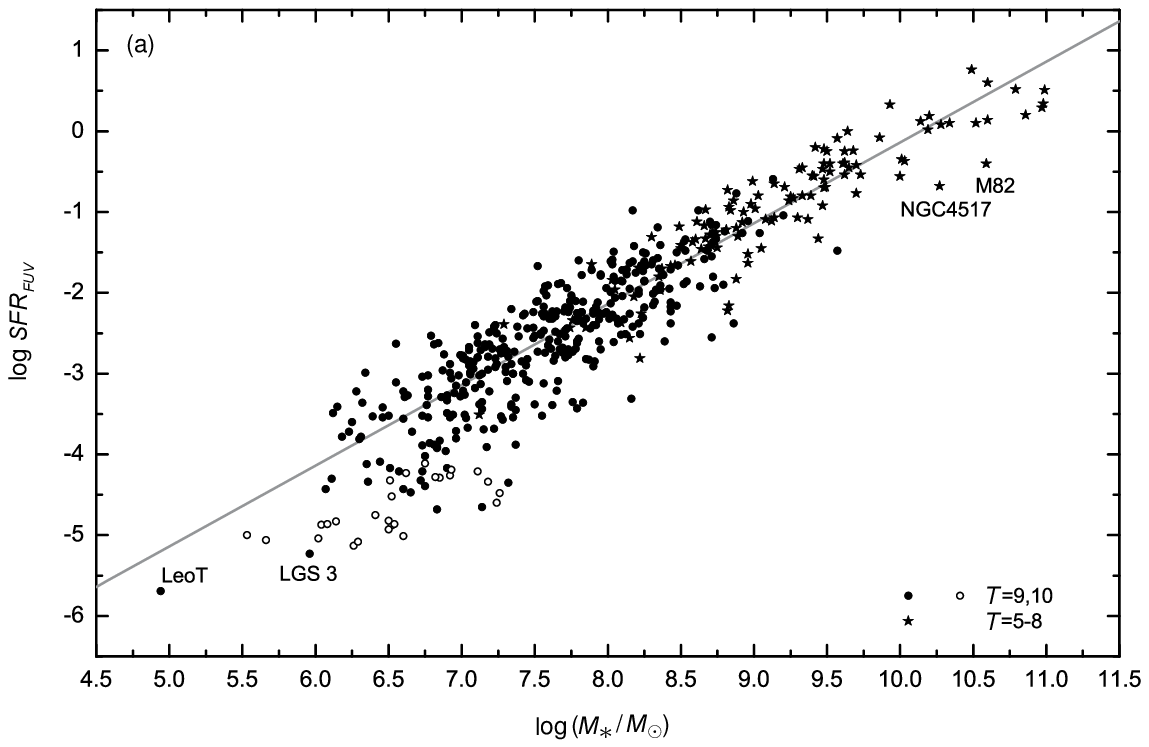}
   \includegraphics[scale=1]{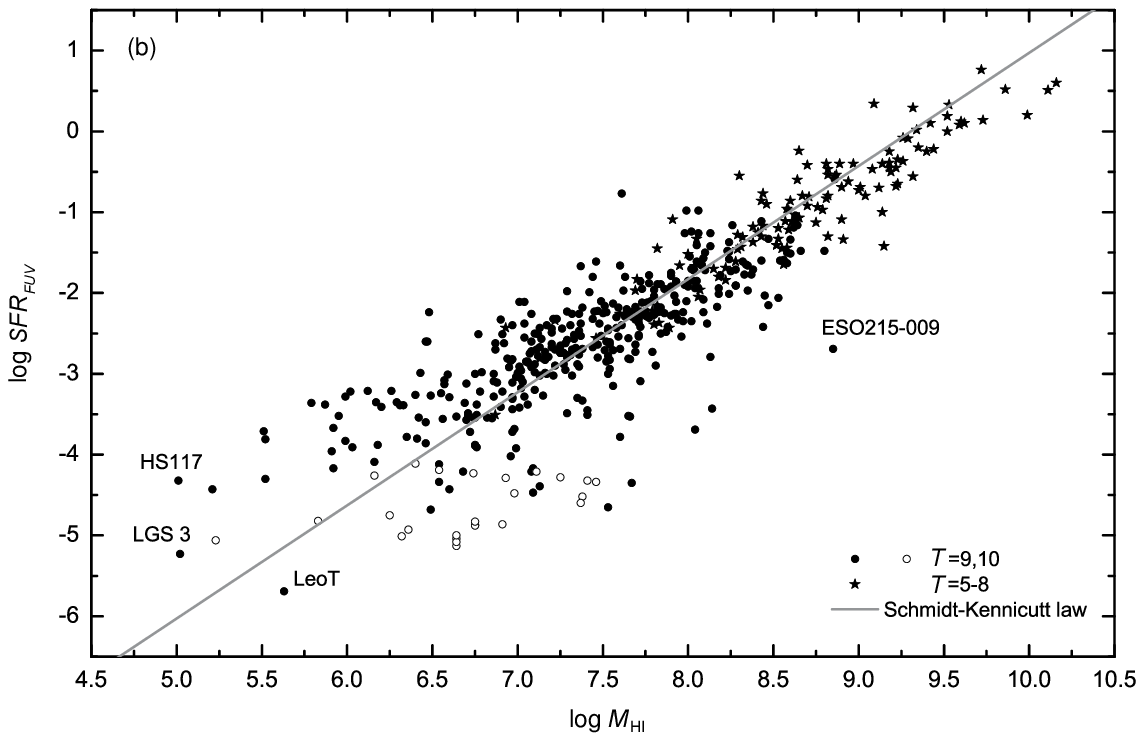}
   \caption{(a)---Distribution of late-type galaxies by star-formation rate and stellar mass.
The straight line corresponds to the Hubble time of 13.7~Gyr
during which the observed stellar mass of the galaxy is reproduced
for the observed rate of star formation. (b)---Distribution of
late-type galaxies by integrated star-formation rate and hydrogen
mass.
   The straight line corresponds to the  Schmidt--Kennicutt relation with a slope of~1.4.}
  \end{figure*}

Figure~7a shows the distribution of late-type dwarf $(T = 9,~10)$
and spiral $(T = 5$--$8)$ galaxies by star-formation rate and
stellar mass. The dwarf and spiral galaxies are shown by circles
and asterisks, respectively. The  cases where only the upper flux
limits were measured are shown by open circles. The dashed line
corresponds to the cosmological time of 13.7~Gyr during which the
galaxy reproduces the observed stellar mass with the observed rate
of star formation. Hence dwarf and spiral galaxies obey the same
relation and the current rate of star formation in most of the
galaxies is quite sufficient to reproduce their stellar mass.

Figure~7b shows the distribution of late-type LV galaxies by
star-formation rate and hydrogen mass. Designations are the same
as in the top panel. The straight line corresponds to the slope of
1.4 known as the Schmidt--Kennicutt relation for individual sites
of star formation in galaxies. The data presented show that
integrated hydrogen mass and integrated rate of star formation of
late-type galaxies follow quite well the dependence established
for sites of star formation located inside them.

In late-type galaxies star-formation rate closely correlates with
stellar mass and therefore specific star-formation rate $sSFR =
SFR/M_*$ referred to unit stellar mass is usually employed to
characterize the star-forming activity of these systems.

\begin{figure*}
\setcaptionmargin{5mm} \onelinecaptionstrue \captionstyle{normal}
\includegraphics[scale=1.2]{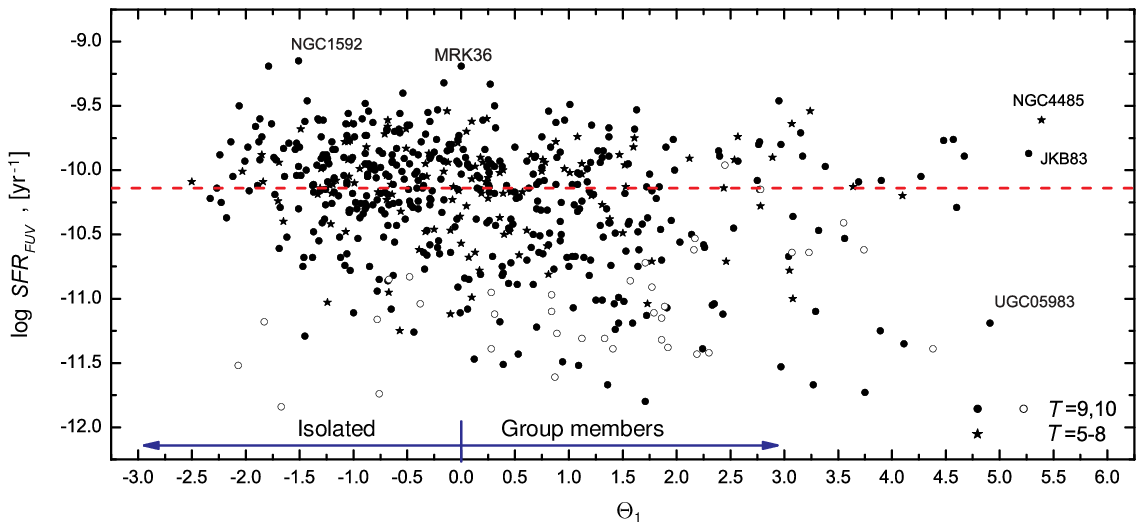}
\caption{Specific star-formation rate and tidal index in late-type
LV galaxies. The horizontal line corresponds to the
\mbox{$sSFR=H_0=13.7$~Gyr$^{-1}$}. Designations are the same as in
Fig.~7.}
  \end{figure*}

Figure~8 shows the distribution of late-type LV galaxies by $sSFR$
and tidal index $\Theta_1$. Galaxies with upper  $FUV$ flux limits
are shown by open circles. Note that parameter $sSFR$ has the
dimension of~yr$^{-1}$, which coincides with that of the Hubble
parameter $H_0=13.7$~Gyr$^{-1}$. Galaxies with $sSFR =H_0$ (the
dashed line) are capable of reproducing their stellar mass with
the observed rate of star formation during the cosmological time
$H_0^{-1}$. It follows from these data that the scatter of $sSFR$
values increases from isolated dwarfs to group members. This
pattern is due to the increase of the fraction of objects in
groups with low star-formation rates because of the likely
``expulsion'' of gas from the dwarf galaxy as it moves through
dense intergalactic medium. Unlike the blurry lower boundary of
the distribution of galaxies by  $sSFR$, the upper boundary
appears rather sharp with the limit $sSFR_{\rm max}\simeq
-9.4$~dex, which depends only slightly on ambient
density~\cite{kar2013b:Karachentsev_n_en,kar2018a:Karachentsev_n_en}. Only a few dwarf galaxies of the
LV (Mrk\,36, NGC\,1592) at the stage of ongoing star formation are
located slightly above this limit. The existence of this limit,
which is similar to the Eddington limit of stellar luminosity, is
indicative of hard feedback of the process: the stronger the
``epidemics'' of star formation in the galaxy, the less resources
of gas remain to sustain it. The value of this limit \mbox
{$sSFR_{\rm max}\simeq 5.5\times H_0$} may serve as an important
characteristic of the process of star formation at the present
epoch $(z=0)$.

It is evident from Fig.~8 that the amplitude of the drop of the
average $sSFR$ from isolated dwarfs to group members is small
compared to the dispersion of the specific rate of star formation.
This indicates that the variety of observed star formation rates
in late-type dwarf galaxies is primarily determined by their
internal parameters and not by the ambient effects.

Measurements of the integrated H$\alpha$-flux of a galaxy can be
used to determine the star-formation rate on a characteristic time
scale on the order of 10~Myr. According to~\cite{ken1998:Karachentsev_n_en},
\begin{equation}
\log SFR=8.98+2\log D+\log F_c({\rm H}\alpha),
\end{equation}
where $SFR$  is in the units of ($M_{\odot}\,{\rm yr}^{-1}$), and
the flux (in erg\,cm$^2$\,s$^{-1}$) is corrected for Galactic and
internal extinction. The most extensive H$\alpha$ survey of LV
galaxies was performed with the 6-m telescope of the Special
Astrophysical Observatory of the Russian Academy of Sciences.
During the last ten years homogeneous \mbox {H$\alpha$-images}
were obtained for more than 400 galaxies (see~\cite{kai2019:Karachentsev_n_en} and
references therein). Combined with the data by other authors,
\mbox {H$\alpha$-images} are currently available for 61\% of LV
galaxies. Among them H$\alpha$ emission was detected in 85\% of
irregular dwarfs,  95\%  of  Im+BCD-type objects, and in 41\% of
spheroidal  (dSph+dE) systems. Curiously, some classical  dSph
dwarfs, e.g.,  DDO\,44, contain small sites of star formation,
which can be seen in H$\alpha$~\cite{kar2011:Karachentsev_n_en}.

\begin{figure*}
\setcaptionmargin{5mm} \onelinecaptionstrue \captionstyle{normal}
\includegraphics[scale=1, angle=90]{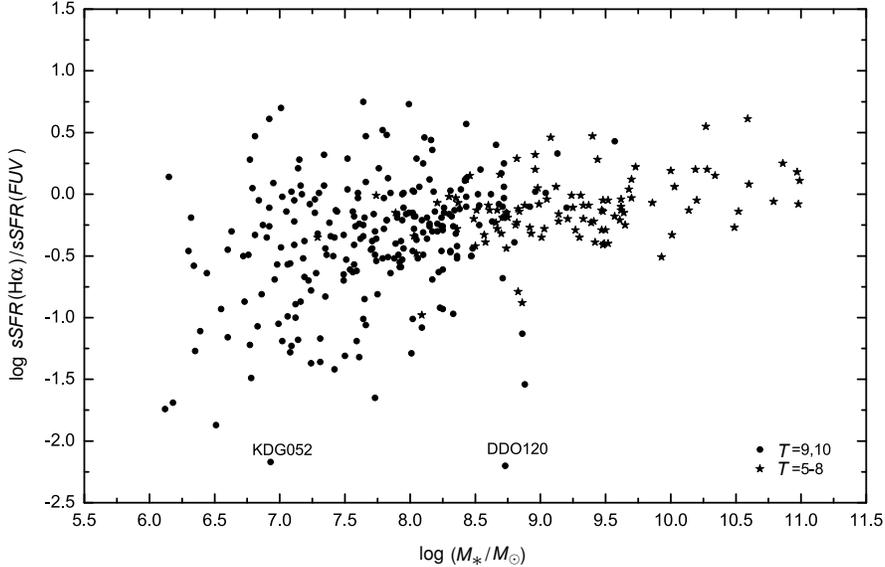}
\caption{The ratio of star-formation rates determined from the
H$\alpha$-line flux to the rates determined from $FUV$-band flux
for late-type LV galaxies. Unlike dwarf galaxies (circles), spiral
galaxies of  Sc--Sm types are shown by asterisks.}
  \end{figure*}

A comparison of the  $SFR$ estimates inferred from H$\alpha$-line
and $FUV$-band fluxes gives the idea of the variations of
star-formation rates in dwarf galaxies on a time scale on the
order of \mbox {10--100}~Myr. Figure~9 reproduces $SFR({\rm
H}\alpha)/SFR(FUV)$---the distribution of the ratio for late-type
galaxies as a function of stellar mass. The above data lead us to
conclude that:
\begin{list}{}{
\setlength\leftmargin{2mm} \setlength\topsep{2mm}
\setlength\parsep{0mm} \setlength\itemsep{2mm} }
 \item (a) Mutual calibration of the empirical  $SFR({\rm H}\alpha)$
and $SFR(FUV)$ values by formulas~(3) and~(4) is quite good for
late-type spiral galaxies, which are dominated by disks.
 \item (b) The $SFR({\rm H}\alpha)/SFR(FUV)$ ratio on the average decreases from spirals to
progressively less massive dwarfs. Various causes of this pattern
were discussed in~\cite{pfl2009:Karachentsev_n_en,wei2012:Karachentsev_n_en,fum2011:Karachentsev_n_en}, however, it is
still not entirely clear why H$\alpha$-flux-determined
star-formation rates in dwarfs are underestimated.
 \item (c) The dispersion of the  $SFR({\rm H}\alpha)/SFR(FUV)$ ratio increases significantly
with decreasing mass of the galaxy. This can be naturally
explained by starburst activity, which becomes relatively more
pronounced in low-mass objects.
 \end{list}
Dependence of the  $SFR({\rm H}\alpha)/SFR(FUV)$ ratio on other
parameters---apparent axial ratio of the galaxy, its morphological
type, ambient density---were analyzed in~\cite{kar2018a:Karachentsev_n_en}.

The evolutionary state of the ensemble of LV dwarf galaxies can be
conveniently characterized by the ``Past--Future''~(PF) diagnostic
diagram where dimensionless parameters
       \begin{equation}
       \begin{array}{rcl}
       P&=&\log(SFR\times T_0/M_*),\\
       F&=& \log(1.85 M_{\rm H\,I}/SFR\times T_0)
       \end{array}
       \end{equation}
do not depend on the errors of the estimated distance to the
galaxy~\cite{kar2007:Karachentsev_n_en,kar2013b:Karachentsev_n_en}.

\begin{figure*}
\setcaptionmargin{5mm} \onelinecaptionstrue \captionstyle{normal}
\includegraphics[scale=1]{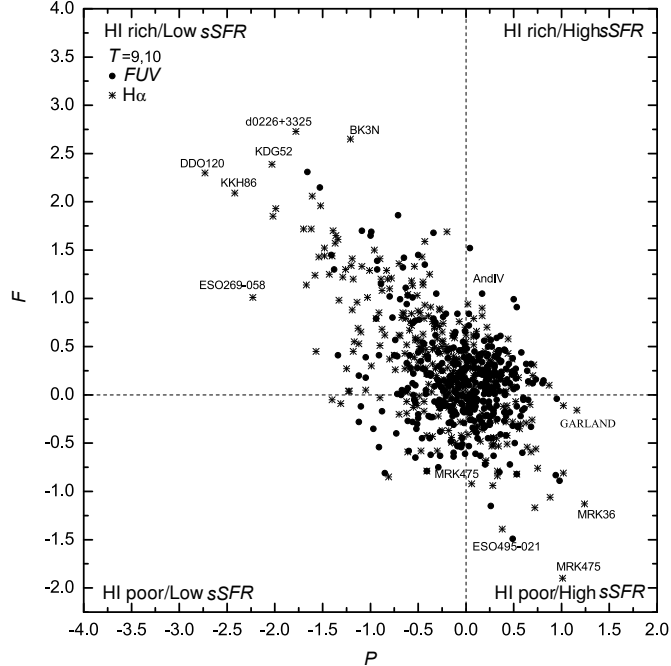}
\caption{The ``Past--Future'' diagnostic diagram for dwarf
galaxies of the types $T = 9,~10$.}
  \end{figure*}

Here parameter $P$ is the specific star-formation rate normalized
to cosmological time $T_0=13.7$~Gyr. Parameter $F$ characterizes
the time (in the units of $T_0$) during which the galaxy exhausts
its available reserves of gas at the observed star-formation rate
$SFR$. The coefficient  1.85 is introduced to take into account
the contribution of  He and $H_2$ to the total mass of
gas~\cite{fuk2004:Karachentsev_n_en}. Note that a galaxy with the coordinates $P=0$,
$F=0$ is capable of reproducing its observed stellar mass during
time $T_0$ and has sufficient reserves of gas to sustain the
observed star-formation rate over yet another time interval $T_0$.

Figure~10 shows the PF diagram for late-type $(T = 9,~10)$ LV
dwarf galaxies. The centroid of the dwarf distribution is located
near the coordinate origin. Hence the typical observed
star-formation rate in dwarfs is quite sufficient for reproducing
their observed stellar masses, and, on the whole, the population
of dwarf galaxies is halfway in processing its gas into stars.

Another important feature of the PF diagram is that the diagonal
extent of the distribution of dwarfs in parameters $P$ and  $F$.
This is evidently due to the starburst activity. During a
starburst dwarf galaxies like Mrk\,36, Mrk\,475 find themselves in
the bottom right part of the PF diagram and as the starburst fades
they move to the top left quadrant where passive dwarfs like
DDO\,120 and KDG\,52 reside. The proportion of dwarfs in quadrants
\mbox {[$P>0$, $F<0$]} and [$P<0$, $F>0$] is of about 3:1, which
indicates that dwarf galaxies spend more time in the quiescent
phase compared to the starburst phase. The presence of ``dead''
dwarfs of  dSph and dE types, which are not represented in Fig.
10, increases this ratio even further.

It is important that such interpretation is based on the
assumption that the galaxy evolves in the  ``closed box'' mode.
The inflow of intergalactic gas observed in some galaxies may have
appreciable effect on the evolutionary tracks of galaxies in the
PF diagram.

\begin{figure*}
\setcaptionmargin{5mm} \onelinecaptionstrue \captionstyle{normal}
\includegraphics[scale=1.2]{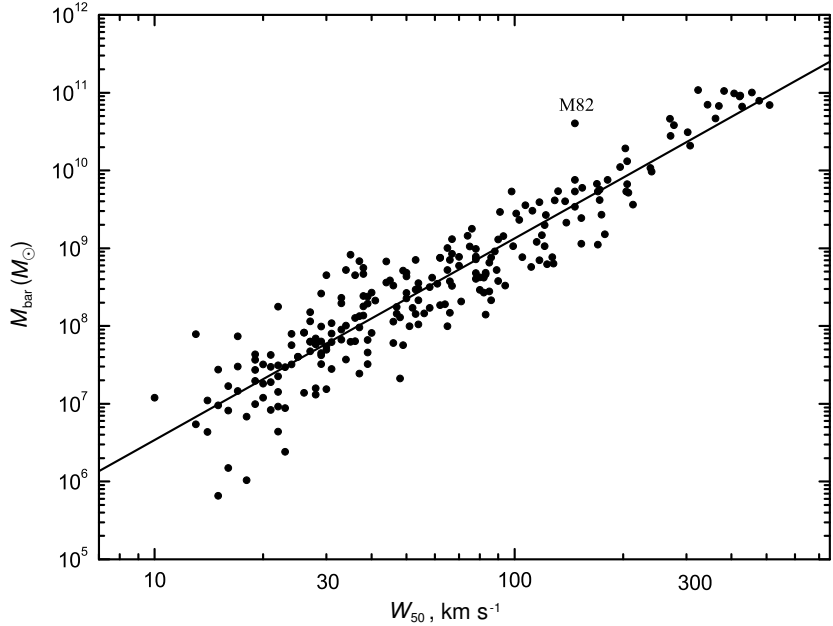}
\caption{Baryonic Tully--Fisher relation for LV galaxies with
accurate distance estimates and inclinations greater than
$45\degr$.}
  \end{figure*}

\section{BARYONIC TULLY--FISHER RELATION}

It follows from the data in Figs.~5 and~6 that about 40\% of
late-type dwarf galaxies in the LV have hydrogen masses greater
than their stellar masses. Given the presence of He and H$_2$,
gas-rich dwarfs are the dominating population of the LV.

The classical Tully--Fisher relation~\cite{tul1977:Karachentsev_n_en} expresses
linear relationship between the logarithm of the width of the
H\,I-line, $W_{50}$, and absolute magnitude (or logarithm of the
stellar mass) of the galaxy. In the domain of dwarf galaxies the
linearity of the  \mbox {TF relation} systematically breaks down
because of the contribution of the gaseous component to the total
baryonic mass of the galaxy. This was first noted by
McGaugh~\cite{mcg2005:Karachentsev_n_en,mcg2012:Karachentsev_n_en}.

The abundance of LV galaxies with accurately measured distances
made it possible to extend the baryonic Tully--Fisher
relation~(bTF) to the domain of low masses down to $\log(M_{\rm
bar}/M_\odot)\simeq6$. According to~\cite{kar2017:Karachentsev_n_en}, the bTF
relation for 330 galaxies in the LV has the form shown in Fig.~11.
The regression line in this figure is described by the following
formula:
\begin{equation}
\log(M_{\rm bar})=2.59\log(W_{50})+3.84,
\end{equation}
where the baryonic mass is in $M_{\odot}$ and the width of the
H\,I at half maximum, in km\,s$^{-1}$. In addition to galaxies
with accurately measured distances we also show in Fig.~11 objects
with distances estimated based on their membership in known
groups.

The standard deviation on the  bTF diagram is equal to 0.39~dex,
making it possible to estimate the distances of other gas-rich
dwarf galaxies with a typical error of 0.195~dex. The main
contributor to this error are  $W_{50}$ measurement errors and
photometric errors. The error of the bTF distance estimates for
galaxies with accurate observational data can be reduced down to
0.10~dex. Further analysis shows~\cite{kar2017:Karachentsev_n_en} that incorporating
a second parameter into equation~(6), such as surface brightness
of the galaxy or the $M_{\rm H\,I}/M_*$, decreases only slightly
the scatter on the diagram.

\section{DWARF GALAXIES AND LOCAL DISTRIBUTION OF DARK MATTER}

The presence of dwarf satellites with accurately measured radial
velocities and distances around nearby massive galaxies made it
possible to determine the total (orbital) masses on the scale
length of virial radius \mbox {$R_{\rm vir}\simeq250$}~kpc. These
estimates for 21 principal LV galaxies are listed above in
Table~1.

Lynden--Bell~\cite{lyn1981:Karachentsev_n_en} proposed an independent method for
determining the total mass of a group (cluster).

A galaxy group as overdensity distorts the Hubble flow of
surrounding galaxies slowing down their velocities. As a result,
the zero-velocity surface forms around the group, which separates
the collapsing region from the zone of cosmic expansion. In the
spherically symmetric case the radius  $R_0$ of the zero-velocity
surface is related to the total mass $M_T$ of the group by the
formula:
\begin{equation}
M_T=(\pi^2/8G) R^3_0 H_0^2/f^2(\Omega_m),
\end{equation}
where dimensionless factor $f$ depends on the average mass density
and varies from 1 to 2/3 with $\Omega_m$ varying from 0 to 1. In
the standard  $\Lambda$CDM model with $\Omega_m=0.24$,
$\Omega_{\lambda}=0.76$, and \mbox
{$H_0=73$}~km\,s$^{-1}$\,Mpc$^{-1}$ from
\cite{spe2007:Karachentsev_n_en} the $M_T(R_0)$ formula acquires
the form:
\begin{equation}
(M_T/M_{\odot})=2.12\times10^{12}(R_0/{\rm Mpc})^3.
\end{equation}

The distribution of velocities and distances of dwarf galaxies
within 3~Mpc of the centroid of the Local group was analyzed
repeatedly by a number of authors~\cite{ekh2001:Karachentsev_n_en,kar2002:Karachentsev_n_en,kar2009:Karachentsev_n_en}.
Figure~12 shows the most complete version  of this diagram
(Fig.~4a from~\cite{kas2018:Karachentsev_n_en}). Non-isolated \mbox {($\Theta_1>0$)}
galaxies burdened with virial velocities are excluded. The Hubble
regression line crosses the zero-velocity line at the level of
\mbox {$R_0\!=\!(0.95\!\pm\!0.03)$}\,Mpc~\cite{kas2018:Karachentsev_n_en}, which
corresponds to the Local group mass of \mbox
{$M_T\!=\!(1.8\!\pm\!0.2)\!\times\!10^{12}\,M_{\odot}$}. This
estimate proved to be  somewhat below the total mass of the Milky
Way and Andromeda galaxy: $M({\rm MW} + {\rm M\,31}) =
2.87\times10^{12}M_{\odot}$.

\begin{figure*}[bpt!!!]
\setcaptionmargin{5mm} \onelinecaptionstrue \captionstyle{normal}
\includegraphics[scale=0.6]{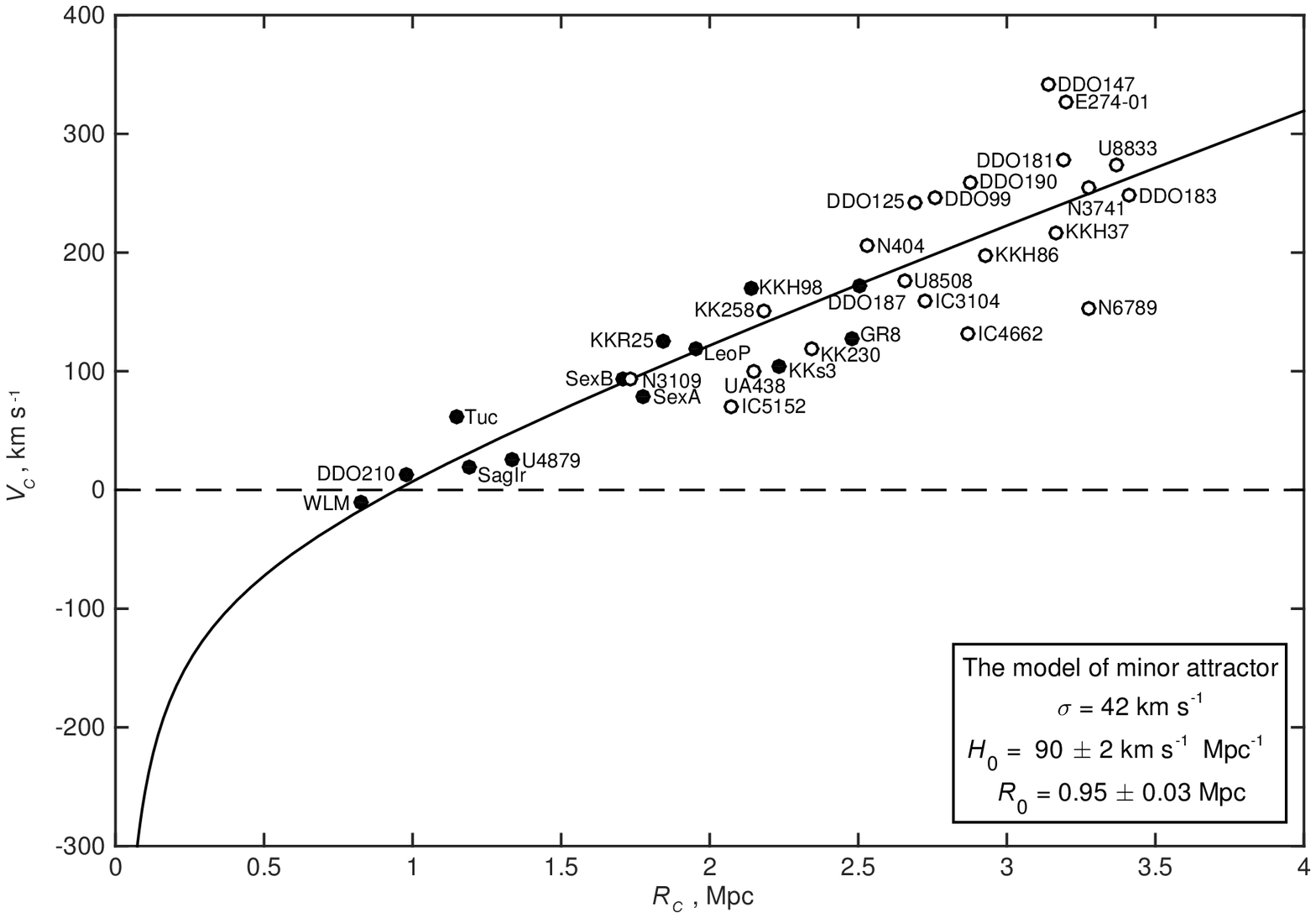}
    \caption{Velocities and distances of the nearest isolated dwarf galaxies relative
    to the Local group barycenter. (Fig.~4a~\cite{kas2018:Karachentsev_n_en}).}
  \end{figure*}

Fairly high accuracy of the $R_0$ and the total mass of the LG
based on external motions of galaxies is due to ``coldness'' of
the Local Hubble flow where the dispersion of radial velocities is
only $\sigma_V=42$~km\,s$^{-1}$\cite{kas2018:Karachentsev_n_en}.

Kashibadze and Karachentsev~\cite{kas2018:Karachentsev_n_en} joined suites of dwarfs
surrounding other dominating galaxies of the LV---M\,81, Cen\,A,
NGC\,4736, M\,101, etc.---to estimate the radius $R_0$ of the
``synthetic'' LV group. The $R_0=(0.92\pm0.02)$~Mpc value
determined from a combined sample of 14 groups corresponds to the
total synthetic group mass of about \mbox
{$1.7\times10^{12}M_{\odot}$}, which is about 60\% of the typical
virial mass of nearby groups. The fact that the estimate of the
total mass based on external motions of galaxies is made on a
scale length that is 3--4 times greater than the virial radius led
Kashibadze and Karachentsev~\cite{kas2018:Karachentsev_n_en} to conclude that there
is no appreciable amount of dark matter in the outskirts of the
Local group and other neighboring groups of galaxies.

A similar approach to determining   the total mass was also
applied to the nearest Virgo cluster of galaxies located at a
distance of 16.5~Mpc. Between the Local group and Virgo cluster
there are about 30 dwarf galaxies with the distances measured
using observations made with the Hubble Space
Telescope~\cite{kar2014d:Karachentsev_n_en,kar2018b:Karachentsev_n_en}.

 \begin{figure*}[bpt!!!]
\setcaptionmargin{5mm} \onelinecaptionstrue \captionstyle{normal}
   \includegraphics[scale=0.7]{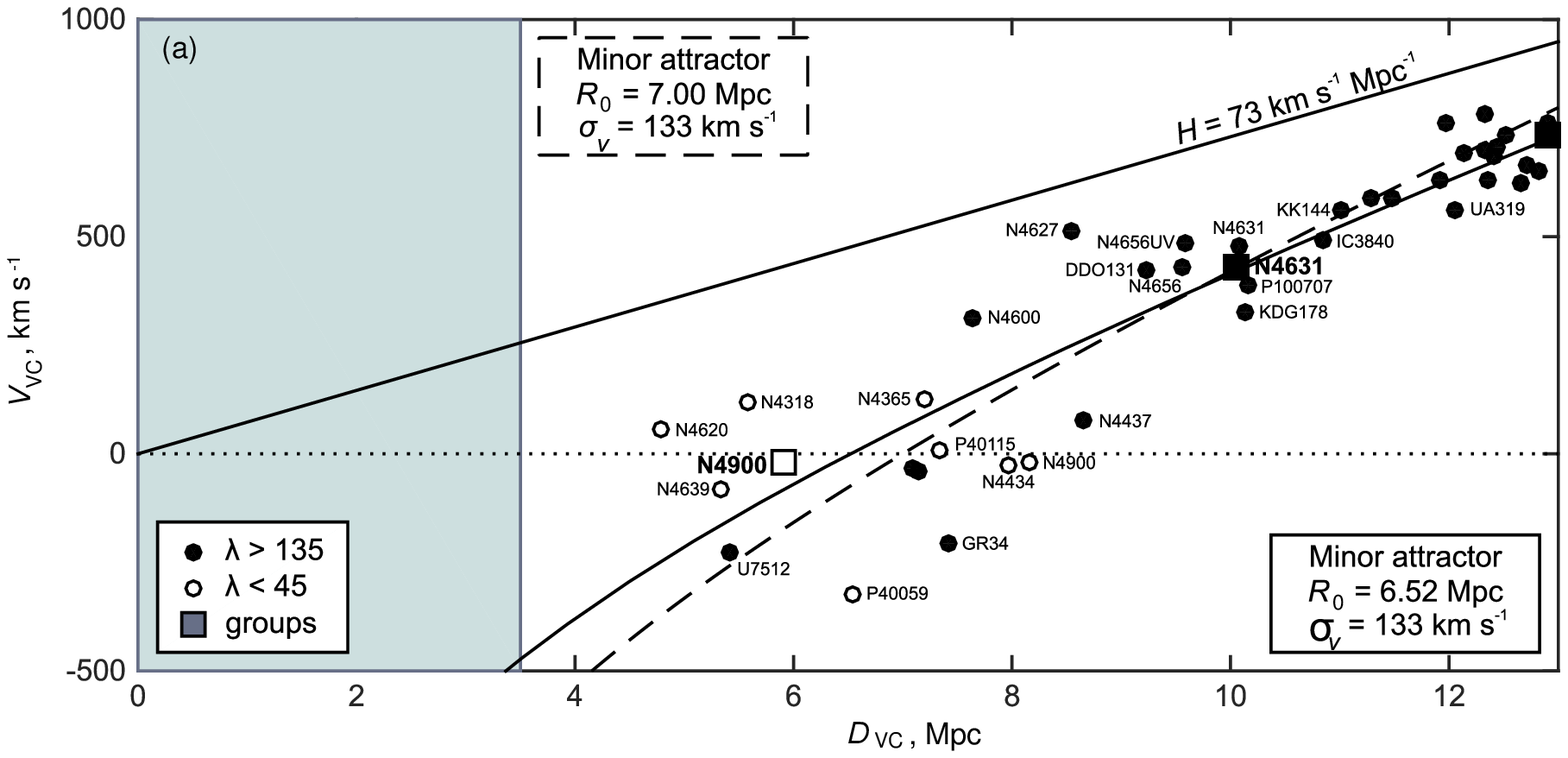}
   \includegraphics[scale=0.7]{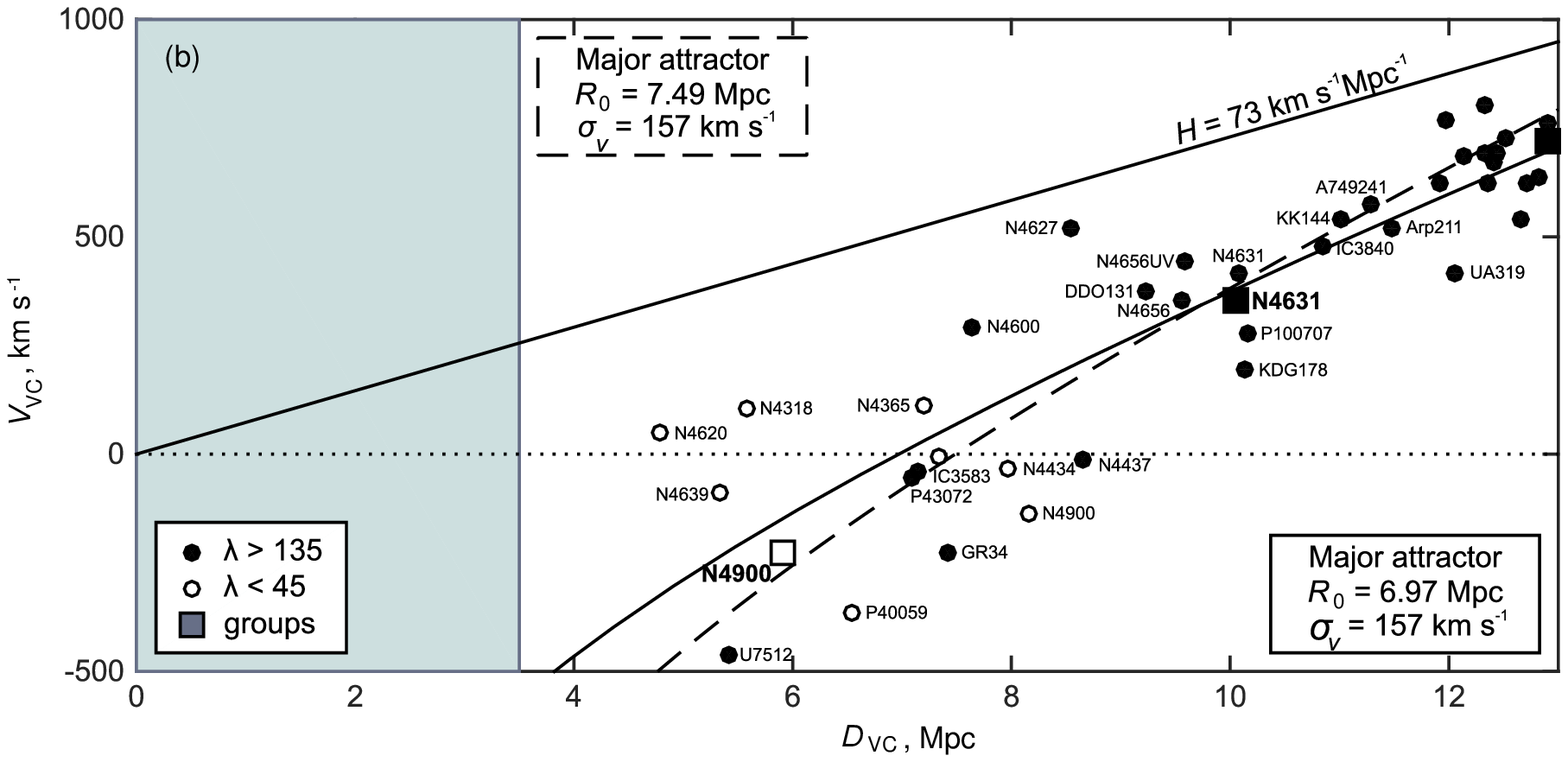}
\caption{Velocities and distances of galaxies located in front of
Virgo cluster relative to  the cluster center estimated under two
different assumptions about the dominant pattern of  motions
(Fig.~13~\cite{kas2018ab:Karachentsev_n_en}).}
  \end{figure*}

The  radius of the LV is equal to 2/3 of the distance to Virgo
cluster. Such an extent of the LV allows tracing the effect of the
fall of galaxies onto the cluster and determine the radius $R_0$
with high accuracy. Figure~13 shows the pattern of the Hubble flow
of galaxies relative to the center of Virgo cluster. Galaxies with
accurately measured velocities are shown by circles. The domain of
virial motions inside the cluster proper is shown by grey color.
The lack of observational data about the tangential components of
galaxy velocities makes it necessary to adopt certain assumptions
concerning their dominating motion pattern. The two panels of the
figure correspond to the extreme assumptions about the pattern of
motions of galaxies near the cluster: the case of a minor
attractor, when Hubble expansion prevails, and major attractor
dominated by radial motions toward the cluster center. According
to these data with the uncertainty of the pattern of galaxy
motions taken into account, the radius of the zero-velocity sphere
around Virgo is \mbox {$R_0=(7.32\pm0.28)$}~Mpc, which corresponds
to the total cluster mass of \mbox
{$M_T=(8.3\pm0.9)\times10^{14}M_{\odot}$}. This estimate agrees
well within the quoted errors with the virial mass of Virgo
cluster $7.5\times10^{14}M_{\odot}$ based on the data adopted
from~\cite{tul1984:Karachentsev_n_en}. Shaya et al.~\cite{sha2017:Karachentsev_n_en} obtained
practically the same estimate of the total mass of Virgo by
numerically modeling galaxy trajectories.

  \begin{figure*}[bpt!!!]
\setcaptionmargin{5mm} \onelinecaptionstrue \captionstyle{normal}
\vbox{
   \includegraphics[scale=0.6]{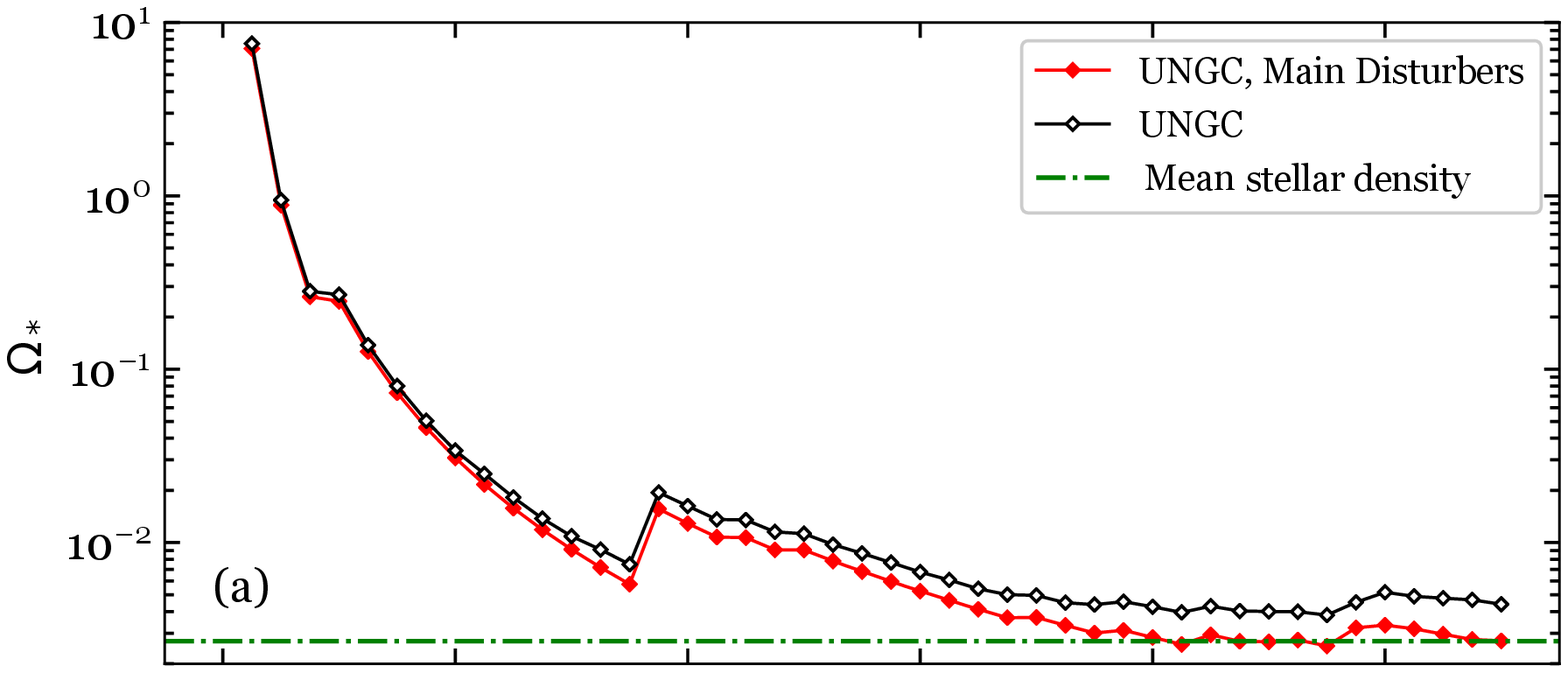}\vspace{2mm}
   \includegraphics[scale=0.6]{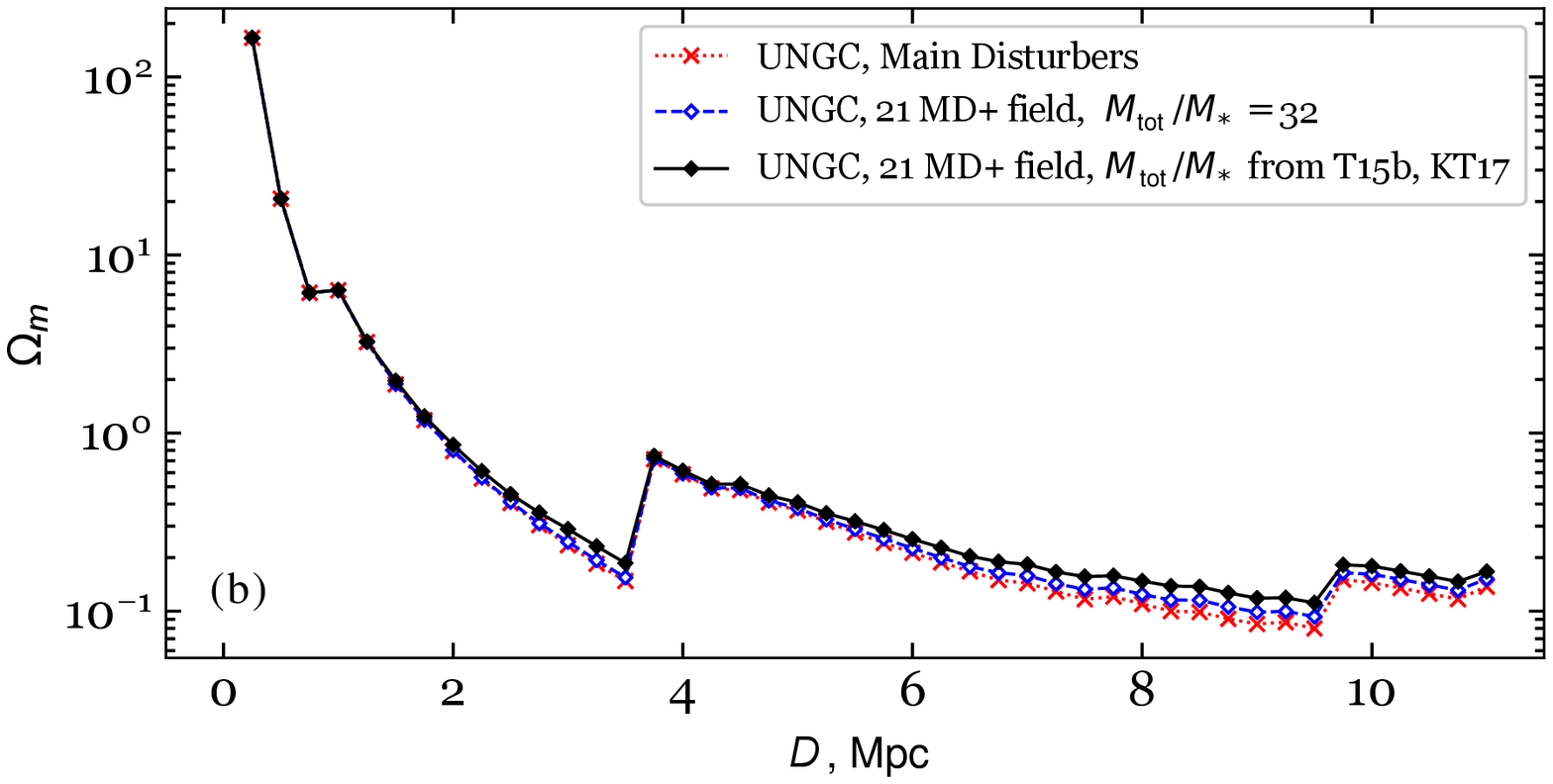}}
\caption{Integrated profiles of stellar and total (virial) mass in
the LV in the units of critical density (Figs.~1 and~2
in~\cite{kar2018c:Karachentsev_n_en}.) }
   \end{figure*}

Karachentsev and Telikova~\cite{kar2018c:Karachentsev_n_en} used the most complete
available data about the distances and halo masses of nearby
galaxies to reconstruct the distribution of visible and dark
matter in the LV. Figure~14a shows the integrated stellar-mass
profile in the units of critical density. The lower line
corresponds to the contribution of the principal LV galaxies
listed in Table~1 and the upper line, the contribution of the
entire LV population to the luminosity density. The horizontal
line fixes the average cosmic density  of stellar mass according
to~\cite{fuk2004:Karachentsev_n_en}. As is evident from the figure, the LV exhibits
excess of stellar density over its global value on all scales out
to the distance of 11~Mpc.

Figure~14b, which shows the distribution of the dark-matter
density, demonstrates a totally different pattern. Here the lower
line shows the contribution of the halos of the principal galaxies
from Table~1, and the middle and upper line correspond to the
total contribution of all galaxies computed assuming (1) that the
$M_T/M_*=32$~\cite{kar2014b:Karachentsev_n_en} ratio is the same for all galaxies or
(2) that $M_T/M_*$ for field galaxies increases toward groups and
dwarf galaxies~\cite{tul2015:Karachentsev_n_en}. Differences between the three lines
are small because the main contributors to the matter density in
the LV are two dozen most massive galaxies. It is remarkable that
the average density of dark matter within 11~Mpc is equal to only
\mbox {$\Omega_m=0.17$}, i.e., the LV is a low-density region in
terms of  this parameter.

\begin{figure*}
\setcaptionmargin{5mm} \onelinecaptionstrue \captionstyle{normal}
\includegraphics[scale=0.65]{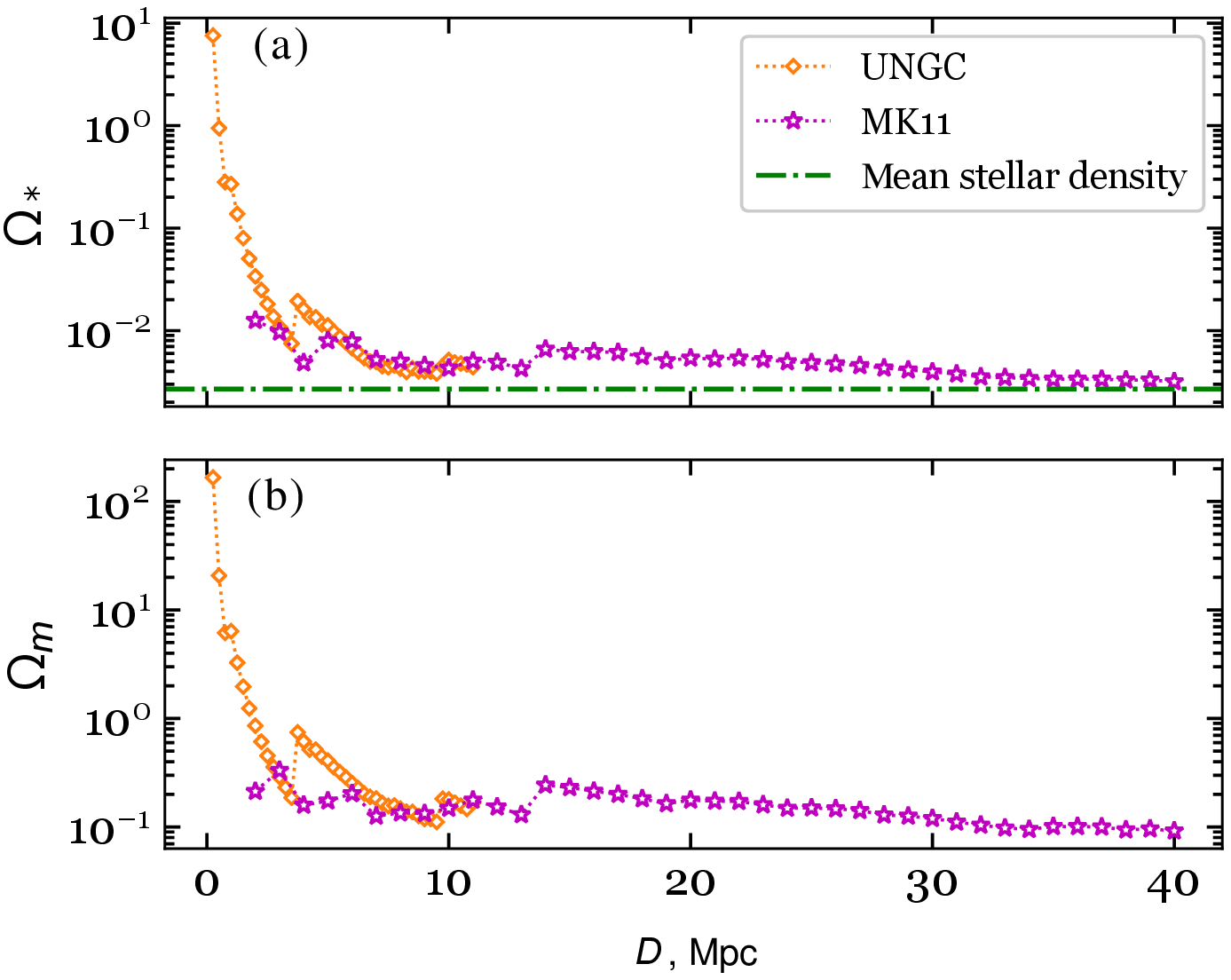}
\caption{Integrated profiles of stellar and dark-matter density
within 40~Mpc of the Milky Way (Fig.~3~\cite{kar2018c:Karachentsev_n_en}.) }
  \end{figure*}

This strange result also shows up in a much greater volume.
Makarov and Karachentsev~\cite{mak2011:Karachentsev_n_en} analyzed the local
distribution of visible and dark (virial) mass based on their
catalog of groups of galaxies. The sequences of asterisks in the
panels of Fig.~15 show the profiles of stellar and virial mass
according to the data of the above authors. The average density
values in the LV are shown by the  diamond signs. It follows from
the data in the top panel that the average local density of
visible matter shows a small excess compared to the global value
out to a distance of 40~Mpc. However, the density of dark matter
proves to be rather low smoothly decreasing down to \mbox
{$\Omega_m=0.09\pm0.03$} at the distance of 40~Mpc. A similar
value, $\Omega_m=0.12$--$0.14$, was recently inferred by Kourkchi
and Tully~\cite{kou2017:Karachentsev_n_en} from the data in a catalog of groups of
galaxies organized in accordance with a different algorithm. The
low value of the local density of dark matter compared to its
global value ($\Omega_m=0.24$--$0.31$) leads one to conclude that
the bulk of the mass of the cosmic dark matter is not concentrated
in groups and clusters but scattered in the intergalactic space
between them.

 \section{CONCLUSIONS}

In recent years comprehensive study of galaxies in the near
Universe crystallized into a separate branch of observational
cosmology. Numerical simulation of the dynamical evolution of
galaxies in terms of the standard  \mbox {$\Lambda$CDM-model} has
reached the spatial resolution on the scale of the sizes and
masses of dwarf galaxies. In the process, the cosmology on small
scales has run into well-known problems: observed deficit of low-
and very low-mass galaxies, observed lack of the central ``cusps''
in the rotation curves of dwarf galaxies, and abundance of thin
flat structures made of dwarf satellites around massive galaxies.
A review of these and other problems of the standard paradigm can
be found in~\cite{kro2016:Karachentsev_n_en}. The problems of standard models, which
became apparent in studies of dwarf galaxies, provide a powerful
stimulus for refining the theoretical scenarios of the formation
and evolution of galaxies.

Precise measurements of the radial velocities and distances to
galaxies make it possible to determine their peculiar velocities
and use these velocities to reconstruct the locations and masses
of surrounding attractors. Numerous precise measurements of galaxy
distances with the Hubble Space Telescope made the LV a unique
laboratory for validating cosmological models as foreseen by
Pebbles~\cite{pee1993:Karachentsev_n_en}.

Where is the bulk of the dark mass located: in systems of galaxies
or in the space between them? So far, we have no a straight answer
to this question. The ever increasing density of data about
peculiar velocities in the LV will serve as the most suitable
observational material for solving this problem.

\begin{acknowledgements}
We are grateful to the staff members of the Laboratory of
Extragalactic Astrophysics and Cosmology of the Special
Astrophysical Observatory of the Russian Academy of Sciences
S.~S.~Kaisin, O.~G.~Kashibadze, G.~G.~Korotkova, D.~I.~Makarov,
L.~N.~Makarova, M.~E.~Sharina, and V.~E.~Karachentseva for their
long-continued assistance in performing this work.
\end{acknowledgements}

\section*{FUNDING}
 This work was supported by the Russian Foundation for Basic Research (grant no.~18--02-00005).

\section*{CONFLICT OF INTEREST}
The authors declare that there is no conflict of interest
regarding the publication of this article.


 {}

\end{document}

%% file: sao_cmd_author.tex
\def\saoname{Special Astrophysical Observatory,  Russian Academy of Sciences,
              Nizhnii Arkhyz, 369167 Russia}

\def\squareforqed{\hbox{\rlap{$\sqcap$}$\sqcup$}}

\def\sq{\ifmmode\squareforqed\else{\unskip\nobreak\hfil
\penalty50\hskip1em\null\nobreak\hfil\squareforqed
\parfillskip=0pt\finalhyphendemerits=0\endgraf}\fi}

\def\degr{\hbox{$^\circ$}}

\def\arcsec{\hbox{$^{\prime\prime}$}}

\def\utw{\smash{\rlap{\lower5pt\hbox{$\sim$}}}}

\def\udtw{\smash{\rlap{\lower6pt\hbox{$\approx$}}}}

\def\fm{\hbox{$\,.\!\!^{\rm m}$}}

\def\diameter{{\ifmmode\mathchoice
{\ooalign{\hfil\hbox{$\displaystyle/$}\hfil\crcr
{\hbox{$\displaystyle\mathchar"20D$}}}}
{\ooalign{\hfil\hbox{$\textstyle/$}\hfil\crcr
{\hbox{$\textstyle\mathchar"20D$}}}}
{\ooalign{\hfil\hbox{$\scriptstyle/$}\hfil\crcr
{\hbox{$\scriptstyle\mathchar"20D$}}}}
{\ooalign{\hfil\hbox{$\scriptscriptstyle/$}\hfil\crcr
{\hbox{$\scriptscriptstyle\mathchar"20D$}}}}
\else{\ooalign{\hfil/\hfil\crcr\mathhexbox20D}}%
\fi}}

\newcommand{\ab}{Astrophysical Bulletin }

\newcommand{\aaa}{Astron. and Astrophys. }

\newcommand{\aas}{Astron. and Astrophys. Suppl. }

\newcommand{\aj}{Astron.~J. }
\newcommand{\apjs}{Astrophys.~J. Suppl. }

\newcommand{\mnras}{Monthly Notices Royal Astron. Soc. }

\newcommand{\alet}{Astronomy Letters }
\newcommand{\an}{Astronomische Nachrichten }

